\documentclass[Crown,sagev,times]{sagej}

\usepackage{moreverb,url}
\usepackage{mathtools}
\usepackage{subcaption}

\usepackage[colorlinks,bookmarksopen,bookmarksnumbered,citecolor=red,urlcolor=red]{hyperref}

\newcommand{\F}{\mathbf{F}}
\newcommand{\B}{\mathbf{B}}
\newcommand{\e}{\mathbf{e}}
\newcommand{\X}{\mathbf{X}}
\newcommand{\x}{\mathbf{x}}
\newcommand{\Chi}{\boldsymbol{\chi}}
\newcommand{\dV}{\Delta}

\DeclareMathOperator{\tr}{tr}

\newcommand\BibTeX{{\rmfamily B\kern-.05em \textsc{i\kern-.025em b}\kern-.08em
T\kern-.1667em\lower.7ex\hbox{E}\kern-.125emX}}

\def\volumeyear{2024}

\begin{document}

\runninghead{Bonavia et al.}

\title{On the Nonlinear Eshelby Inclusion Problem and its Isomorphic Growth Limit}

\author{J. E. Bonavia\affilnum{1}, S. Chockalingam\affilnum{3,4} and T. Cohen\affilnum{1,2}}

\affiliation{\affilnum{1}Massachusetts Institute of Technology, Department of Mechanical Engineering, Cambridge MA, USA\\
\affilnum{2}Massachusetts Institute of Technology, Department of Civil and Environmental Engineering, Cambridge MA, USA\\
\affilnum{3}Massachusetts Institute of Technology, Department of Aeronautical and Astronautical Engineering, Cambridge MA, USA\\
\affilnum{4}Purdue University, Department of Mechanical Engineering, West Lafayette IN, USA}

\corrauth{Tal Cohen, \\
Massachusetts Institute of Technology,
Department of Civil and Environmental Engineering, \\
77 Massachusetts Avenue, 
Cambridge MA,
02139, USA.}

\email{talco@mit.edu}

\begin{abstract}
In the late 1950's, Eshelby's linear solutions for the deformation field inside an ellipsoidal inclusion and, subsequently, the infinite matrix in which it is embedded were published. The solutions' ability to capture the behavior of an orthotropically symmetric shaped inclusion made it invaluable in efforts to understand the behavior of defects within, and the micromechanics of, metals and other stiff materials throughout the rest of the 20\textsuperscript{th} century. Over half a century later, we wish to understand the analogous effects of microstructure on the behavior of soft materials; both organic and synthetic; but in order to do so, we must venture beyond the linear limit, far into the nonlinear regime. However, no solutions to these analogous problems currently exist for non-spherical inclusions. In this work, we present an accurate semi-inverse solution for the elastic field in an isotropically growing spheroidal inclusion embedded in an infinite matrix, both made of the same incompressible neo-Hookean material. We also investigate the behavior of such an inclusion as it grows infinitely large, demonstrating the existence of a non-spherical asymptotic shape and an associated asymptotic pressure. We call this the isomorphic limit, and the associated pressure the isomorphic pressure. 

\end{abstract}


\maketitle
\section{Introduction}
Many of the most pressing problems in the physics of soft solids are inextricably linked to the physics of incompatibility.\cite{residual} Growing biological objects such as tumors or biofilms are often constrained within, or against, other solid masses.\cite{cheng2009,mills2014,zhang2021} So much of the behavior of biological materials is dependent on the micromechanical incompatibilities and interactions between the cells and other microstructures that exist at a scale above the molecular but beneath our own.\cite{plants,skin,holzapfel_gasser} Even in synthetic materials, differences in rates of curing can lead to incompatibility driven stress concentrations within critical components and the resulting deformations can be large. Understanding these concentrations is especially important as we lean more heavily into the use of resins and other polymers as engineering materials.\cite{PCPdefects1,PCPdefects2,PhotoCuredEshelby}

The problem of an elastic inclusion embedded in an infinite matrix, as first formulated by John D. Eshelby in 1957, has served as a fundamental and instrumental example of incompatibility in the physics of solids.\cite{eshelby1957,eshelby1959} Eshelby set out to model any number of physical problems where \textit{``...the uniformity of an} [infinite] \textit{elastic medium is disturbed by a region within it which has changed its form...''}. He called the changing region the inclusion and the rest of the infinite body the matrix. Ultimately he was able to solve this problem for an arbitrarily shaped inclusion and to find a particularly elegant solution for the case where this region took on the shape of an ellipsoid. \textit{``Fortunately''}, remarks Eshelby in 1957, \textit{``the general ellipsoid is versatile enough to cover a wide variety of particular cases.''}

The power and versatility of Eshelby's solution is two-fold: its ability to handle the case of an inclusion of an orthotropically symmetric shape and the fact that it is fully analytical. As it turns out, Eshelby was quite right. Over the years, a truly staggering number of examples were able to be well approximated as ellipsoidal inclusions (as they are convex, vaguely round, and longer in some directions than others). His exact solution has been instrumental in understanding the fields which develop around defects in stiff materials like metal, and through homogenization methods (which utilized the analytical nature of the solution) has laid the foundation for the development of micromechanical models for metals and stiff composite materials.\cite{mura,moritanakacomposites,composites} Extensions of his 1957 and 1959 works to other shapes and materials have been applied to better understand a wide variety of materials including metamaterials and piezoelectric materials. \cite{metamaterials,piezo1997,SharmaPiezo, liping}

One might imagine that in trying to understand the physics of incompatibility in soft solids, Eshelby's solution to the inclusion problem may serve as a strong basis in doing so. However, although exact, Eshelby's solution, and the thought experiment he utilized to attain it, rely on the geometric and constitutive linearity inherent to small deformations. Unfortunately, in the soft systems we wish to understand in the 21\textsuperscript{st} century, we do not have the luxury of this linearity.

Generally speaking, exact analytical solutions to nonlinear problems, especially those lacking strong symmetries, are sparse at best. The inclusion problem is no exception. The extent to which exact solutions to the nonlinear inclusion problem have been found (largely by Yavari\cite{yavarianisotropic}, Golgoon\cite{Golgoon_Yavari_2017}, and Goriely\cite{yandgisotropic}) for commonly used constitutive relations is limited to specific spherical and circular cylindrical inclusions changing their shapes radially, axisymmetrically, torsionally, or azimuthally such that they remain spherical or circular cylindrical in their deformed shape. The problem of an elliptical cavity or rigid inclusion (both limits of the problems known as inhomogeneities) have also been solved  in two dimensions using a specific two-dimensional constitutive relation called a harmonic material (whose physical applications are greatly limited). \cite{HarmonicVoid2024, HarmonicStiff2024} However, no exact solutions exist for a general nonlinear ellipsoidal inclusion (or inhomogeneity) inside an infinite matrix. 

This has led many to turn to numerical methods such as finite element analysis (FEA) as a first line solution to approximately solve nonlinear problems. For instance, in 2000, Diani and Parks\cite{parks} examined the effect of geometric nonlinearity on the ellipsoidal inclusion problem by using FEA to solve the problem with a logarithmic strain energy function. However, there are a few downsides to using these numerical methods, especially for a problem like the inclusion problem. Eshelby's original statement of the problem imagines that the inclusion is embedded within an \emph{infinite} matrix. When solving the problem numerically one cannot have a truly infinite matrix. As the matrix is made larger and larger it should converge to the solution of an infinite matrix, however, this also greatly increases the computational cost of the simulation, often taking hours of simulation time to reach moderate levels of stretch, and days to weeks of simulation time when performing parameter sweeps. Additionally, because a numerical solution is not expressed analytically, it is difficult to prove any sort of asymptotic behavior of the system and any homogenized solutions tend to be even more resource intensive.

Another set of methods that can be used to find approximate (or exact) solutions to nonlinear problems are inverse, or semi-inverse methods. In an inverse solution, a fully determined, kinematically-admissible deformation field is assumed and then the resulting stress field is checked to see if it satisfies the necessary field and boundary equations. In a semi-inverse solution (as first developed by St. Venant), a form for the deformation field is assumed involving some set of unknowns, and the unknowns are then used to satisfy the relevant field equations and boundary conditions of the system. If these equations can be exactly satisfied then you have found the exact solution to the problem. However, even if the equations are not exactly satisfied, the closest approximate solution which satisfies the initial kinematic assumptions can be found by minimizing the total elastic energy of the system. \cite{phdthesisSemiInverse} Inverse and semi-inverse methods have been used readily in finding fairly accurate approximate solutions to problems involving spherical and spheroidal voids in nonlinear elastic and plastic materials.\cite{HouAbeyaratne, GologanuProlate, GologanuOblate, PorousEllipsoids, li2022, PhotoCuredEshelby}

In this work, we utilize a semi-inverse method to formulate an approximate solution to the problem of a nonlinear elastic, incompressible, isotropically growing ellipsoidal inclusion embedded within an infinite matrix made of the same material. We begin by examining Eshelby's solution in order to formulate a key assumption about the ``shape'' of the nonlinear deformation: that there exists a set of ellipsoids which fully cover the infinite body, and which remain ellipsoids after deformation. We then use these kinematic assumptions, along with incompressibility, to simplify our description of the field. We find that the entire field can be described by two functions which describe this set of ellipsoids. We then return to Eshelby's solution to select this set of ellipsoids, generalizing such that we retain two ``knobs'' which we can tune to minimize the total elastic energy of the system, a la the semi-inverse method. 

We apply this formulation to the problems of an isotropically growing spheroidal inclusion and demonstrate the striking accuracy of the resulting solutions through comparison with FEM simulations and Eshelby's solution deep into the nonlinear range. Finally, we analyze the behavior of this solution as the inclusion grows infinitely large and show that for the case of spheroidal inclusions there exists a non-spherical shape which the inclusion tends towards. We call this limit the isomorphic growth limit and demonstrate its properties. We find that in the isomorphic growth limit the pressure inside the inclusion approaches a limit higher than that of spherical cavitation and call this pressure limit the isomorphic pressure. We end the work with a discussion of the application of this solution to homogenization, and opine on the implications of the ``shape'' of the resulting fields for this approximate nonlinear solution to a canonical problem.

\section{The Inclusion Problem in Finite Elasticity}

\begin{figure}[t]
    \centering
    \includegraphics[width=0.98\textwidth]{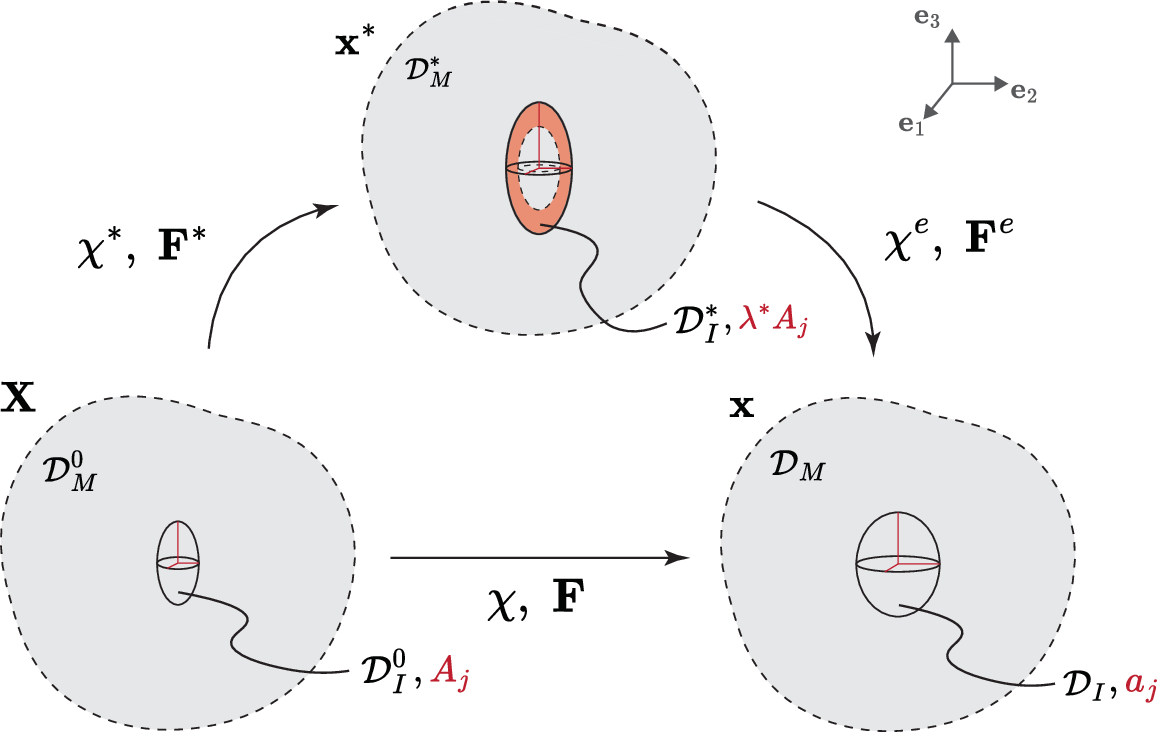}
    \caption{The infinitely large body $\mathcal{B}$ in the reference (bottom left), transformed (top center), and deformed (bottom right) configurations. In each configuration the region occupied by the inclusion (denoted by the subscript ``$I$'') and matrix (denoted by the subscript ``$M$'') are indicated. The inclusion has the three semi-axes $A_j$ in the reference configuration, $\lambda^*A_j$ in the transformed configuration, and $a_j$ in the deformed configuration aligned with the right-handed Cartesian coordinate system ($\e_1,\e_2,\e_3$). The red region in the transformed  configuration denotes the incompatibility present between the inclusion and matrix. The mappings $\Chi^e$, and $\Chi$ and their gradients $\F^*$, $\F^e$, and $\F$ represent the transformation, elastic part of the deformation, and total deformation respectively. } 
    \label{fig:PotatoDiagram}
\end{figure}

Consider an infinitely large body, $\mathcal{B}$, described by a set of material points  $\X\in \mathbb{R}^3$.    We divide this body into two  material sub-regions; the \textit{inclusion} - $\mathcal{B}_I$,
and the \textit{matrix} within which it is enclosed - $\mathcal{B}_M$, such that $\mathcal{B}_I\cup\mathcal{B}_M\equiv \mathcal{B}$ and $\mathcal{B}_I\cap\mathcal{B}_M$ forms a closed and simply connected surface we call the \emph{inclusion-matrix interface}. 

Initially, in their stress-free state, the inclusion and the matrix occupy the respective regions $\mathcal{D}^0_I$ and $\mathcal{D}^0_M$ (Fig. \ref{fig:PotatoDiagram}), and  the inclusion-matrix interface is denoted by $\partial\mathcal{D}^0_I$.
The inclusion  then undergoes a permanent, inelastic transformation, such as growth or a phase transition, which without the constraint imposed on it by the surrounding matrix, would cause an arbitrary homogeneous transformation stretch. 
However, since the matrix and inclusion constrain each other, they both deform to accommodate the transformation.  The corresponding regions occupied by the deformed inclusion and matrix are denoted by $\mathcal{D}_I$ and $\mathcal{D}_M$, respectively, and the inclusion-matrix interface is denoted by  $\partial\mathcal{D}_I$. Material points in the deformed configuration are thus mapped to their new, deformed coordinates $\x = \Chi(\X)$.

In the context of finite elasticity we restrict our attention to homogeneous, isotropic, incompressible, and hyperelastic materials and assume that the matrix and inclusion are perfectly bonded at their interface, namely that the deformation is continuous ($\Chi(\X)\in C^0$). Under these assumptions, we seek a solution to the deformation field for a given transformation stretch.

\bigskip

Or, in the words of Eshelby, we now task ourselves with solving for the \textit{``... elastic state of inclusion and matrix."}

\subsection{Kinematic Framework}
We define the deformation gradient as 
\begin{equation}\label{eq:defF}
    \F = \frac{\partial \Chi(\X)}{\partial \X}\;.
\end{equation}
However, because the transformation of the inclusion alters its stress-free configuration, the elastic free energy of the solid will no longer be dependent on $\F$, but instead on only the part of the deformation which forces the two bodies together elastically. 

In order to isolate this part of the deformation we introduce a motion $\x^* = \Chi^*(\X)$ we call the \textit{transformation} which maps material point $\X$ to the point $\x^*$ which it occupies in the transformed body $\mathcal{D}^*$. In this transformed configuration the inclusion occupies the region $\mathcal{D}^*_I$, the matrix occupies the region $\mathcal{D}^*_M$ (as shown at the top of Fig. \ref{fig:PotatoDiagram}). We note that the transformed configuration is \emph{incompatible}, meaning that gaps and/or overlaps exist between $\mathcal{D}^*_I$ and $\mathcal{D}^*_M$ (as denoted by the shaded red region in Fig. \ref{fig:PotatoDiagram}). This implies that $\Chi^*$ is not a bijective map.

We now decompose $\Chi$ into the transformation $\Chi^*$ and the elastic deformation $\x=\Chi^e(\x^*)$ as $\x = \Chi^e (\Chi^*(\X))$, separating the motion into two steps. The first transforms the inclusion's stress free configuration, and the second deforms the bodies to fit together. Consequently we can decompose the deformation gradient as
\begin{equation}\label{eq:FDecomp}
    \F=\F^e\F^*\;,
\end{equation}
where $\F^e = \partial\Chi^e/\partial\x^*$ is the elastic part of the deformation gradient and $\F^* = \partial\Chi^*/\partial\X$ is the transformation gradient.\footnote{Alhasadi and Federico\cite{Decomp} considered several possible decompositions of the nonlinear version of Eshelby's inclusion problem. We arrived at one of the simplest; that used by Diani and Parks\cite{parks}. }

Because both the inclusion and matrix are made of an elastically incompressible material, we may restrict ourselves to isochoric deformations. This implies that the elastic part of the deformation must obey the constraint
\begin{equation}\label{eq:IncompInt}
    \int_\Omega(J^e-1)\mathrm{dv} = 0\;,
\end{equation}
for any arbitrary sub-body in the matrix or inclusion, where $J^e = \det(\F^e)$. Since (\ref{eq:IncompInt}) must hold for any arbitrary sub-body $\Omega$, the integrand of (\ref{eq:IncompInt}) vanishes everywhere, implying $J^e = 1\;, \forall\; \X$. Further, from (\ref{eq:FDecomp}), we find that incompressibility implies
\begin{equation}\label{eq:JDecomp}
    J = J^*\;, \;\forall\;\X\;,
\end{equation}
where $J = \det(\F)$ and $J^* = \det(\F^*)$. We note that because we assume that the transformation is uniform and occurs only inside the inclusion, $\F^*$ and $J^*$ are piecewise constant with $\F^* = \mathbf{1}$ and $J^*=1$ for all $\X\in\mathcal{D}^0_M$, where $\mathbf{1}$ denotes the identity tensor.

\section{Ellipsoidal Inclusions Undergoing Isotropic Transformations}
Having described the general finite elastic inclusion problem, we now restrict our attention to  ellipsoidal inclusions which undergo isotropic transformations. The semi-axes of an inclusion are taken as $(A_1, A_2, A_3)$, which align with the directions of a right handed Cartesian system  $(\e_1,  \e_2, \e_3)$, respectively, and we set the origin of the coordinate system at the geometric center of the inclusion. 
We describe the body, $\mathcal{B}$, as a continuous set of ellipsoidal surfaces where the dimensionless radial parameter  $\Lambda=\bar{\Lambda}(\X)$ is constant on a given ellipsoid and the functions $\Phi_{2}=\hat{\Phi}_{2}(\Lambda)$ and $\Phi_{3}=\hat{\Phi}_{3}(\Lambda)$ are the corresponding aspect ratios, to write the implicit relationship \begin{equation}\label{eq:RefEllipsoids}
    X_1^2 + \frac{X_2^2}{\Phi_{2}^2} + \frac{X_3^2}{\Phi_{3}^2} = A_1^2\Lambda^2\;,
\end{equation}
where $X_j\equiv\X\cdot\e_j$. According to this definition, $\Lambda$ corresponds to the normalized distance from the origin to the intercept of the ellipsoid with $\e_1$, such that $\Lambda=X_1/A_1$. The inclusion-matrix interface is thus at $\Lambda=1$ and we enforce that this set of ellipsoidal surfaces includes the inclusion-matrix interface by setting $\hat{\Phi}_{2}(1) = \Phi_{02} = A_2/A_1$ and $\hat{\Phi}_{3}(1) = \Phi_{03} = A_3/A_1$. Consequently, \begin{equation} \label{eq:RegionDef}
    \text{for} \quad
    \begin{cases}
    0\leq\Lambda\leq 1, & \X\in\mathcal{D}^0_I  \\
     \Lambda= 1, & \X\in\partial\mathcal{D}^0_I \\
    \Lambda\geq1, & \X\in \mathcal{D}^0_M \;.
    \end{cases}
\end{equation}
Provided the above definitions, we write the  volume of the inclusion as
\begin{equation}\label{eq:VolumeInclusion}
    V^0 = \frac{4\pi}{3}A_1A_2A_3 = \frac{4\pi}{3}\Phi_{02}\Phi_{03}A_1^3\;.
\end{equation}

In the following sections we use the summation convention of Mura\cite{mura}, which operates by the following rules:
\begin{enumerate} 
    \item Repeated lowercase indices are summed from 1 to 3.
    \item Upper case indices take on the same value as the corresponding lower case indices but are not summed. 
\end{enumerate}
Without loss of generality, we also define $\Phi_1=\hat{\Phi}_{1}(\Lambda)\equiv 1$, and consequently $\hat{\Phi}_{1}(1) = \Phi_{01} = A_1/A_1 = 1$, for use in calculations using this summation convention. 

\subsection{The Linear Limit and the Confocal Ellipsoids}
Before attempting to solve the nonlinear inclusion problem, it is instructive to re-examine the small-strains result. In his linear solution, Eshelby \cite{eshelby1957, eshelby1959} defines the inclusion problem in terms of a stress-free infinitesimal transformation strain (eigenstrain), related to $\F^*$ in the above described nonlinear formulation as
\begin{equation}\label{eq:estardef}
    \boldsymbol{\epsilon}^* = \frac{1}{2}\left[\F^*+(\F^*)^T\right]-\mathbf{1}\;.
\end{equation}
He showed that in the linear limit, when $\boldsymbol{\epsilon}^*$ (and by extension $\F^*$) is constant for all $\X\in\mathcal{D}^0_I$, the deformation is uniform inside the inclusion. This implies that in the linear limit $\F$ should be constant inside the inclusion and the inclusion should remain ellipsoidal.  

From Eshelby's solution, as formulated in \cite{eshelby1957,eshelby1959,mura}, for the case of an isotropic transformation strain of the form
\begin{equation}\label{eq:isoTStrain}
    \boldsymbol{\epsilon}^* = \epsilon^*\mathbf{1}\;, \;\;\forall\;\X\in\mathcal{D}^0_I\;,
\end{equation}
it can be trivially shown that if we choose the set of ellipses confocal with $\partial\mathcal{D}^0_I$
\begin{equation}\label{eq:PhiCj}
    \Phi_{J} = \hat{\Phi}^C_J(\Lambda)=\frac{\sqrt{\Phi_{0J}^2+(\Lambda^2-1)}}{\Lambda},\;\;\forall\;\X\in\mathcal{D}^0_M\;,
\end{equation}
then the deformation takes the form
\begin{equation}\label{eq:Xscale}
    x_j = \hat{f}_J(\Lambda)X_j\;,\;\forall\;\X\in\mathcal{D}^0_M\;,
\end{equation}
where $x_j\equiv\x\cdot\e_j$ is the cartesian coordinate of $\x$ in the $\e_j$ direction and the functions $\hat{f}_J$ are determined via the equations of linear elasticity. The above equation implies that for isotropic transformation strains, each of these confocal ellipsoidal surfaces is transformed into a new ellipsoidal surface in the deformed configuration. \footnote{The set of ellipsoids described by $\hat{\Phi}^C_J$ can also be thought of as the ellipsoids which make up an ellipsoidal coordinate system. A thorough derivation of the formulas in Section 3.1 is given in \cite{myThesis, PhotoCuredEshelby}.}

\subsection{Matching The Linear Limit: Ellipsoids to Ellipsoids}
From (\ref{eq:estardef}) and (\ref{eq:isoTStrain}) we find that an isotropic transformation strain corresponds to an isotropic transformation gradient of the form:
\begin{equation}\label{eq:IsotropicFStar}
    \F^* = \lambda^*\mathbf{1}\;, \;\forall\;\X\in\mathcal{D}^0_I\;,
\end{equation}
where $\lambda^*=1+\epsilon^*$ is the transformation stretch. Notice also that this implies $J^*=J^*_I\equiv(\lambda^*)^3$ for all $\X\in\mathcal{D}^0_I$. Since Eshelby's solution is \emph{exact} in the linear limit, the kinematics of any  exact solution to the nonlinear problem of  an ellipsoidal inclusion under an isotropic tranformation \emph{must} reduce to ellipsoids which are transformed to ellipsoids in the linear limit.

Therefore, in order to find an approximate solution to the nonlinear ellipsoidal inclusion problem under an isotropic transformation we choose a kinematically admissable set of deformations in which ellipsoids of constant $\Lambda$, including $\partial\mathcal{D}^0_I$, are transformed to a new set of ellipsoids via the mapping
\begin{equation}\label{eq:xofX}
    \x = \Chi(\X) = \left\{ \frac{\alpha}{\Lambda}X_1\:,\; \frac{\varphi_{2}\alpha}{\Phi_{2}\Lambda}X_2\:,\; \frac{\varphi_{3}\alpha}{\Phi_{3}\Lambda}X_3\right\}\;,
\end{equation} 
where we define the field variables $\alpha =\hat{\alpha}(\Lambda)$, $\varphi_{2}=\hat{\varphi}_{2}(\Lambda)$ and $\varphi_{3}=\hat{\varphi}_{3}(\Lambda)$ as the deformed dimensionless radial parameter and deformed aspect ratios respectively. \footnote{A similar assumption of spherical or ellipsoidal shells remaining ellipsoidal in the deformed configuration is used by Hou and Abeyaratne\cite{HouAbeyaratne}, for spheroidal shells by Gologanu et al \cite{GologanuProlate, GologanuOblate} and by Avaz and Naghdabadi\cite{PorousEllipsoids}. The latter two works assume an ellipsoidal coordinate system and thus implicitly choose the confocal set of ellipsoids. Additionally in \cite{myThesis,PhotoCuredEshelby} the looser assumption of ellipsoids remaining ellipsoids is employed.} The semi-axes of the deformed inclusion are taken as $(a_1, a_2, a_3)$, which align with   $(\e_1,  \e_2, \e_3)$, respectively, and the origin of the coordinate system remains at the geometric center of the inclusion. 

As such, $\alpha(1)=a_1/A_1$, $\hat{\varphi}_{2}(1) = \varphi_{02} \equiv a_2/a_1$, and $\hat{\varphi}_{3}(1) = \varphi_{03} \equiv a_3/a_1$ respectively since this set of ellipsoidal surfaces includes the inclusion-matrix interface (with aspect ratios $\varphi_{02}$ and $\varphi_{03}$). Note that we do not yet enforce a specific form of $\Phi_{j}$, only the values of the undeformed aspect ratios at the matrix-inclusion interface. The volume of $\mathcal{D}_I$ is given as
\begin{equation}\label{eq:VolumeDefInclusion}
    V^* = \frac{4\pi}{3}a_1a_2a_3 = J^*_I V^0\;,
\end{equation}
since the inclusion has grown isotropically by a factor of $\lambda^*$ and the elastic part of the deformation is isochoric. We also define $\varphi_1=\hat{\varphi}_{1}(\Lambda)\equiv 1$, and consequently $\hat{\varphi}_{1}(1) = \varphi_{01} \equiv a_1/a_1 = 1$ for use in calculations using summation convention. In this way
\begin{equation}\label{eq:xofXSum}
    x_j = \frac{\varphi_J}{\Phi_J}\frac{\alpha}{\Lambda}X_j\;.
\end{equation}

By inserting (\ref{eq:xofXSum}) into (\ref{eq:defF}), and remembering that $\Lambda$ is a function of $\X$, we find that $F_{jk} \equiv \F:(\e_j\otimes\e_k)$
is
\begin{equation}\label{eq:FjkUnsimp}
    F_{jk} = \frac{\partial x_j}{\partial X_k} = \frac{\varphi_J}{\Phi_J}\left(\bar{\lambda}\delta_{jk} + \frac{(\lambda+\beta_J)-\bar{\lambda}(1+\gamma_J)}{A^2\Lambda^2\Gamma}\cdot\frac{X_jX_k}{\Phi_K^2}\right),
\end{equation}
where $\delta_{ij}$ is the Kronecker delta, 
\begin{equation}\label{eq:gammasum}
    \Gamma = 1+\gamma_N\frac{X_nX_n}{\Phi_N^2}\;,
\end{equation}
$\bar{\lambda} = \alpha/\Lambda$, $\lambda = \alpha'$, $\beta_J = \varphi_J'\alpha/\varphi_J$, and $\gamma_J = \Phi_J'\Lambda/\Phi_J$ with $\bullet'$ denoting differntiation with respect to $\Lambda$. We can now find $J$ by taking the determinant of $\F$ and substitute $\Phi_1=\varphi_1=1$. Setting $J=J^*$ equal to $J^*_I$ inside the inclusion and $1$ inside the matrix yields
\begin{equation}\label{eq:J3Darb}
    J = \frac{\varphi_{2}\varphi_{3}}{\Phi_{2}\Phi_{3}}\bar{\lambda}^2\left(\frac{\lambda+\beta_2\frac{X_2^2}{\Phi_2^2}+\beta_3\frac{X_3^2}{\Phi_3^2}}{1+\gamma_2\frac{X_2^2}{\Phi_2^2}+\gamma_3\frac{X_3^2}{\Phi_3^2}}\right) =
    \begin{cases}
        J^*_I, & \forall\;\X\in\mathcal{D}^0_I\\
        1, & \forall\;\X\in\mathcal{D}^0_M\\
    \end{cases}.
\end{equation}
Since (\ref{eq:JDecomp}) must hold for all $\X\in\mathcal{D}^0$ (arbitratry choice of $X_2$ and $X_3$) we find three constraints due to incompressibility 
\begin{subequations}
    \begin{equation}\label{eq:IncompRadialGen}
        \frac{\varphi_2\varphi_3}{\Phi_2\Phi_3}\bar{\lambda}^2\lambda = 
        \begin{cases}
        J^*_I, & \forall\;\X\in\mathcal{D}^0_I\\
        1, & \forall\;\X\in\mathcal{D}^0_M\\
        \end{cases}\;,
    \end{equation}    
    \begin{equation}\label{eq:IncompGam2Gen}
        \beta_2=\lambda\gamma_2\;,
    \end{equation}
    \begin{equation}\label{eq:IncompGam3Gen}
        \beta_3=\lambda\gamma_3\;.
    \end{equation}    
\end{subequations}
For ease of calculations it is also useful to employ the conservation of volume of each ellipsoidal subregion of $\mathcal{B}$ as
\begin{equation}\label{eq:TotIncomp}
    \varphi_2\varphi_3\alpha^3 = 
    \begin{cases}
        J^*_I\Phi_2\Phi_3\Lambda^3\;, & \forall\;\X\in\mathcal{D}^0_I\\
        \left(\Phi_2\Phi_3\Lambda^3+\dV\right)\;, & \forall\;\X\in\mathcal{D}^0_M\\
    \end{cases}
\end{equation}   
where 
\begin{equation}
    \dV = \frac{3}{4\pi A_1^3}(V^*-V^0)  = \Phi_{02}\Phi_{03}(J^*_I-1).
\end{equation}

Inserting (\ref{eq:IncompRadialGen}) into (\ref{eq:TotIncomp}) for all $\X\in\mathcal{D}^0_I$ yields the integrable ordinary differential equation $\alpha'=\alpha/\Lambda$ and thus we find
\begin{equation}\label{eq:GenUnifAlpha}
    \hat{\alpha}(\Lambda)=C_1\Lambda\;, \;\;\;\;\forall\;\X\in\mathcal{D}^0_I\;,
\end{equation}
where $C_1$ is an arbitrary integration constant. This implies that the deformation must be uniform inside the inclusion in order to satisfy incompressibility, a result which matches Eshelby's solution in the linear limit. Because the deformation is uniform within the inclusion, any choice of $\hat{\Phi}_J$ ($\hat{\Phi}_J(1)=\Phi_{0J}$) is kinematically admissible, and thus we choose the convenient $\Phi_J=\Phi_{0J}$ and $\varphi_J=\varphi_{0J}$ for all $\X\in\mathcal{D}^0_I$. In this way from (\ref{eq:TotIncomp}), for the inclusion we find
\begin{equation}\label{eq:UnifAlpha}
    \hat{\alpha}(\Lambda)=\lambda^*\left(\frac{\Phi_{02}\Phi_{03}}{\varphi_{02}\varphi_{03}}\right)^{\frac{1}{3}}\Lambda\;, \;\;\;\;\forall\;\X\in\mathcal{D}^0_I\;.
\end{equation}
Inserting (\ref{eq:UnifAlpha}) into (\ref{eq:FjkUnsimp}) and then applying (\ref{eq:FDecomp}) yields the elastic deformation gradient
\begin{equation}\label{eq:FjkSimpInc}
    F^e_{jk} = \frac{\varphi_{0J}}{\Phi_{0J}}\left(\frac{\Phi_{02}\Phi_{03}}{\varphi_{02}\varphi_{03}}\right)^{\frac{1}{3}}\delta_{jk}\;, \;\;\;\;\forall\;\X\in\mathcal{D}^0_I\;.
\end{equation}

For the matrix, inserting (\ref{eq:IncompRadialGen}) into (\ref{eq:TotIncomp}) and integrating yields
\begin{equation}\label{eq:alphaMatGen}
    \ln(\alpha)=\int{\frac{\Phi_2\Phi_3\Lambda^2}{\dV+\Phi_2\Phi_3\Lambda^3}\;\mathrm{d}\Lambda}\;, \;\;\;\;\forall\;\X\in\mathcal{D}^0_M\;,
\end{equation}
and inserting (\ref{eq:IncompGam2Gen}) and (\ref{eq:IncompGam3Gen}) into (\ref{eq:FjkUnsimp}) yields
\begin{equation}\label{eq:FjkSimp}
    F^e_{jk} = \frac{\varphi_J}{\Phi_J}\left(\bar{\lambda}\delta_{jk} + \frac{(\lambda-\bar{\lambda})(1+\gamma_J)}{A^2\Lambda^2\Gamma}\frac{X_jX_k}{\Phi_K^2}\right)\;, \;\;\;\;\forall\;\X\in\mathcal{D}^0_M\;.
\end{equation}
We have now reduced the problem to the determination of the two undeformed aspect ratio functions $\hat\Phi_2(\Lambda)$ and $\hat\Phi_3(\Lambda)$. We now task ourselves with determining forms of these functions which accurately approximate the true solution.

\subsection{A Generalized Set of Ellipses}

\begin{figure}[t]
    \centering
    \includegraphics[width=0.95\textwidth]{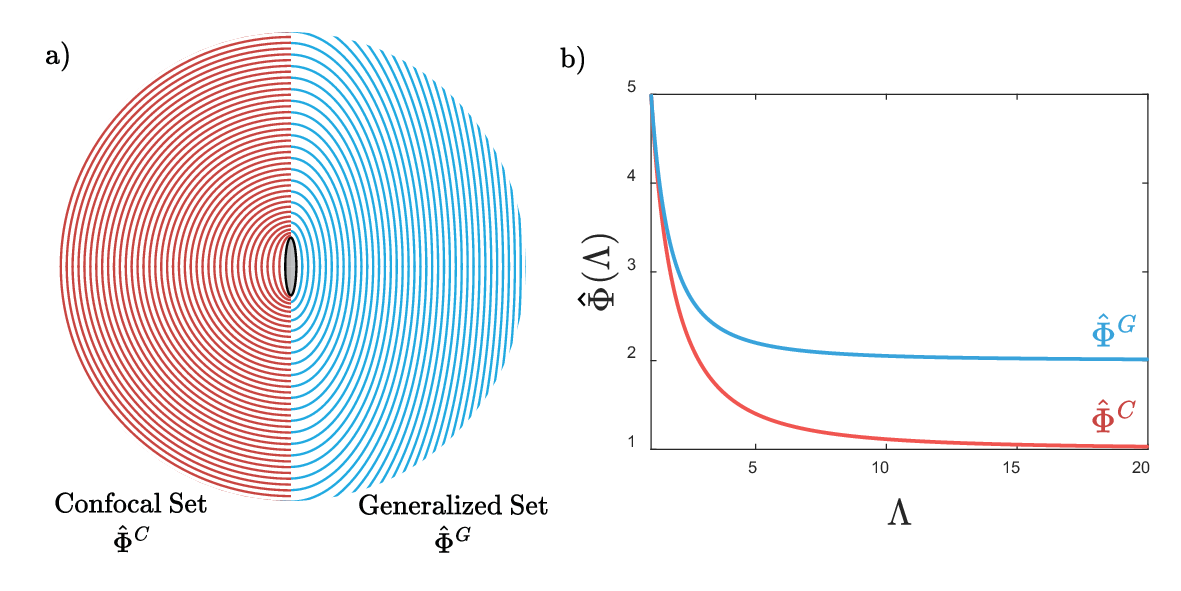}
    \caption{(a) A visualization of the difference between a confocal set of ellipses, characterized by the aspect ratio function $\hat\Phi^C(\Lambda)$, and the generalized set of ellipses, characterized by the aspect ratio function $\hat\Phi^G(\Lambda)$ with $\Phi_\infty=2$. The gray region represents an undeformed inclusion with $\Phi_0=5$. (b) The aspect ratio functions $\hat\Phi^C(\Lambda)$ and $\hat\Phi^G(\Lambda)$ versus $\Lambda$ of (a).}
    \label{fig:Comparison}
\end{figure}

In the following sections we will apply this general framework for ellipsoids to the specific case of spheroidal inclusions ($A_1=A_2$) growing isotropically. We will begin choosing a set of confocal ellipsoids ($\Phi_J=\hat{\Phi}_J^C$). In this case the deformation will be fully constrained by incompressibility and we can find a fully analytical set of expressions that define the deformation. Although a deformation of the form in (\ref{eq:PhiCj}) has been used in the past, we will show that it is a poor approximation of the true solution at even moderate levels of volume growth. Instead, we will show that generalizing $\Phi_J$ to the form
\begin{equation}\label{eq:PhiGj}
    \Phi_{J} = \hat{\Phi}^G_J(\Lambda)=\frac{\sqrt{\Phi_{0J}^2+\Phi_{\infty J}^2(\Lambda^2-1)}}{\Lambda},\;\;\;\;\forall\;\X\in\mathcal{D}^0_M\;,
\end{equation}
leads to a more accurate solution, especially at large deformations. 
 This generalized form describes a set non-confocal ellipsoids (formed by scaling the confocal ellipses uniformly, but anisotropically) which include the inclusion-matrix interface, where the parameters $\Phi_{\infty J}$ describes the limiting value of their aspect ratios in the far-field ($\Lambda\to \infty$). We note that $\hat\Phi_J^G(\Lambda)$ is a monotonic function for all $\Phi_{0J}$,$\Phi_{\infty J}$. 
 An illustration of the differences between the aspect ratio functions $\hat\Phi^C(\Lambda)$ and $\hat\Phi^G(\Lambda)$ is shown in Fig. \ref{fig:Comparison}. In all cases, to remain consistent with (\ref{eq:RefEllipsoids}), we will take $\Phi_{\infty 1}=1$. This set of ellipsoids reduces to the confocal set of ellipsoids in the case where $\Phi_{\infty J}=1$. Thus, we establish that this generalization can properly match the exact solution in the linear limit. For ease of calculations, it can also be shown from (\ref{eq:PhiGj}) that for these generalized ellipsoids
\begin{equation}\label{eq:gammaSub}
    \gamma_J = \frac{\Phi_{\infty J}^2}{\Phi_{J}^2}-1\;.
\end{equation} 

We have now further reduced the problem to the determination of the two far field aspect ratio \emph{parameters} $\Phi_{\infty 2}$ and $\Phi_{\infty 3}$. However, $\Phi_{\infty 2}$ and $\Phi_{\infty 3}$ cannot be determined by incompressibility, and so we turn to energy minimization to find the most accurate approximate solution to the problem which satisfies our kinematic assumptions. 

\subsection{The Elastic Free Energy and Energy Minimization}
We begin by defining an elastic free energy per unit transformed volume $\bar{\psi}^*(\F^e)$. For an incompressible hyperelastic medium, the free energy may be expressed as functions of the first and second invariants of the elastic left Cauchy-Green deformation tensor
$\B=\F^e(\F^e)^T$, as $\hat{\psi}^*(I_1,I_2)$, where $I_1=\tr(\B)=F^e_{jk}F^e_{jk}$ and $I_2 = (\tr(\B)^2-\tr(\B^2))/2$. For this problem however, it is most convenient to use the free energy per unit reference volume (since it can be integrated over the continuous $\mathcal{D}^0$), which can be expressed as
\begin{equation}\label{eq:FERV}
    \hat{\psi}^0(I_1,I_2,J^*) = J^*\left(\hat{\psi}^*(I_1,I_2)\right)\;,
\end{equation}
since $J^*$ represents the ratio of the volume of an element in the transformed configuration to its volume in the reference configuration. In this work we will limit our analysis to a neo-Hookean finite-elastic medium with an elastic free energy density function of the form
\begin{equation}\label{eq:NHFE}
    \hat{\psi}^*(I_1) = \frac{\mu}{2}(I_1-3)\;,
\end{equation}
where $\mu$ is the ground state shear modulus of the material. \footnote{The ground state shear modulus of the inclusion $\mu_I$ and the matrix $\mu_M$ need not be the same (we imagine in this case that the transformation has also changed the material properties of the inclusion), however we consider the transformed sub-body is called the \emph{inclusion} only in the case of $\mu_I=\mu_M$, otherwise the transformed body is known more broadly as an \emph{inhomogeneous inclusion}. The main focus of this work is to find an approximate solution to the specific problem of an isotropically growing ellipsoidal inclusion.} The total elastic free energy in the system for a neo-Hookean inclusion-matrix system can thus be given from (\ref{eq:FERV}) and (\ref{eq:NHFE}), remembering that $J^*=1$ for all $\X\in\mathcal{D}_M^0$, as
\begin{equation} \label{eq:totalFreeEnergyArb}
    \Psi = \int_{\mathcal{D}^0}\psi^0\;\mathrm{dv}^0  =\overbrace{\frac{\mu}{2}\int_{\mathcal{D}^0_M}(I_1-3)\;\mathrm{dv}^0}^{\displaystyle\Psi_M} + \overbrace{J^*_I\frac{\mu}{2}\int_{\mathcal{D}^0_I}(I_1-3)\;\mathrm{dv}^0}^{\displaystyle\Psi_I}\;,
\end{equation}
where $\Psi_M$ and $\Psi_I$ are the total elastic free energy in the inclusion and matrix respectively. 

Choosing $\Phi_J=\hat\Phi^G_J(\Lambda)$ and non-dimensionalizing (\ref{eq:totalFreeEnergyArb}) yields a non-dimensional elastic free energy of the form
\begin{equation}\label{eq:psibarDef}
    \frac{2\Psi}{\mu V^0} = \bar{\Psi}(\lambda^*,\Phi_{0 2},\Phi_{\infty 2},\Phi_{0 3},\Phi_{\infty 3}) = \bar{\Psi}_M + \bar{\Psi}_I\;.
\end{equation} 
where $\bar{\Psi}_M$ and $\bar{\Psi}_I$ are the non-dimensional elastic free energies of the matrix and inclusion respectively. The closest approximate solution for an ellipsoidal inclusion of initial aspect ratios $\Phi_{0J}$ with an isotropic eigenstretch $\lambda^*$, satisfying our kinematic assumptions, can thus be found via the energy minimization conditions
\begin{equation}\label{eq:minimizationConditions}
    \frac{\partial}{\partial\Phi_{\infty 2}}\bar{\Psi}(\lambda^*,\Phi_{0 2},\Phi_{\infty 2},\Phi_{0 3},\Phi_{\infty 3})=0\;\;\;\mathrm{and}\;\;\;\frac{\partial}{\partial\Phi_{\infty 3}}\bar{\Psi}(\lambda^*,\Phi_{0 2},\Phi_{\infty 2},\Phi_{0 3},\Phi_{\infty 3})=0\;.
\end{equation} 

\subsection{Stress State Inside the Inclusion}
For an incompressible neo-Hookean material, the Cauchy stress tensor $\mathbf{T}$ can be calculated as
\begin{equation}\label{eq:CauchyNH}
    \mathbf{T} = \mu\mathbf{B}^d - p\mathbf{1}\;,
\end{equation}
where $\boldsymbol\cdot^d$ represents the deviatoric part of a tensor, defined as $\mathbf{A}^d \equiv\mathbf{A}-(\tr(\mathbf{A})/3)\cdot\mathbf{1}$, and $p$ is an arbitrary hydrostatic pressure field associated with the incompressibility constraint. 
From (\ref{eq:GenUnifAlpha}) we can see that, given our kinematic assumptions, the deformation is uniform inside the inclusion. This, along with mechanical equilibrium ($\mathrm{div}(\mathbf{T}) = \mathbf{0}$) also implies that the stress $\mathbf{T}$ is constant inside the inclusion and can be determined by solving for the spatially constant inclusion pressure $p = p_I$.

By equating the change in internal energy of the matrix with the mechanical work done on it by the applied tractions at the inclusion-matrix interface, we can write the following configurational force balance: 
\begin{equation}\label{eq:GPowerStatement}
    \frac{\partial\Psi_M}{\partial J^*_I} = -\int_{\partial\mathcal{D}_I}\left(\mathbf{T}\mathbf{n}\cdot\frac{\partial\mathbf{x}}{\partial J^*_I}\right)\mathrm{da}\;.
\end{equation}
Remembering that $p=p_I$ is constant along $\partial\mathcal{D}_I$ and noticing that from mass conservation
\begin{equation}
\displaystyle\int_{\partial\mathcal{D}_I}\left(\mathbf{n}\cdot\frac{\partial\mathbf{x}}{\partial J^*_I}\right)\mathrm{da}=\int_{\mathcal{D}_I}\mathrm{div}\left(\frac{\partial\mathbf{x}}{\partial J^*_I}\right)\mathrm{dv}=\frac{\partial V^*}{\partial J^*_I} = V^0,
\end{equation}
inserting (\ref{eq:CauchyNH}) into (\ref{eq:GPowerStatement}) yields
\begin{equation}\label{eq:genqEq}
    p_I = \frac{1}{V^0}\left[{\frac{\partial\Psi_M}{\partial J^*_I}+\mu\int_{\partial\mathcal{D}_I}\left(\B^d\mathbf{n}\cdot\frac{\partial\mathbf{x}}{\partial J^*_I}\right)\mathrm{da}}\right]\;.
\end{equation}

\subsection{Spheroids ($A_1=A_2$)}
The following is a summary of the equations which can be used to solve for the field in a spheroidal inclusion and the surrounding matrix during isotropic transformations, given our kinematic assumptions and choosing the generalized version of the undeformed aspect ratio function (\ref{eq:PhiGj}). Detailed derivations of these equations can be found in Appendix A. 

For the case of a spheroid, from axisymmety around $\e_3$, we assume that $\hat{\Phi}_2(\Lambda)=\hat{\Phi}_1(\Lambda)\equiv 1$ and $\hat{\varphi}_2(\Lambda)=\hat{\varphi}_1(\Lambda)\equiv 1$. As such we are left with a single aspect ratio $\Phi_3$ in the reference configuration, and $\varphi_3$ in the deformed configuration. For simplicity in the section on spheroids we will drop the subscript ``3'' in the following section. That is to say that $(\Phi_3,\Phi_{03},\Phi_{\infty 3},\varphi_3, \varphi_{03}, \gamma_3)\equiv(\Phi,\Phi_{0},\Phi_{\infty},\varphi, \varphi_{0}, \gamma)$.

The deformation $\Chi(\X)$ can be described via the following functions for the undeformed aspect ratio 
\begin{subequations}
    \begin{equation}\label{eq:PhiMatSpheroid}
        \hat\Phi(\Lambda)=
        \begin{cases}
            \Phi_{0}, & 0<\Lambda< 1 \\
             \displaystyle\frac{\sqrt{(\Phi_{0}^2-\Phi_{\infty }^2)+\Phi_{\infty }^2\Lambda^2}}{\Lambda}, & \Lambda>1\;,
        \end{cases}
    \end{equation}
 and the corresponding deformed radial parameter $\alpha$ reads
    \begin{equation}\label{eq:alphaMatSpheroid}
        \hat\alpha(\Lambda)=
        \begin{dcases}
            \lambda^*\left(\frac{\Phi_{0}}{\varphi_{0}}\right)^{\frac{1}{3}}\Lambda, & 0<\Lambda< 1 \\
            \frac{1}{\Phi_\infty}\prod_{R}(\Phi\Lambda-R)^{\frac{R^2}{3R^2-\delta}}, & \Lambda>1\;,
        \end{dcases}
    \end{equation}
where $\delta = (\Phi_0^2-\Phi_\infty^2)$ and each $R$ is a root of the depressed cubic polynomial $(R^3-\delta R + \Phi_\infty^2\dV = 0)$. Note that we have not decided a priori whether or not the spheroidal inclusion is prolate ($A_3>A_1=A_2$) or oblate ($A_3<A_1=A_2$). Our formulation is agnostic to this choice, changing only the number of real $R$ values. In either case, $\alpha$ will be entirely real. In the rest of this section we find it convenient to define $\alpha_0\equiv\hat{\alpha}(1)$ and to use the relation
\begin{equation}\label{eq:JStarSpheroid}
J^*_I = (\lambda^*)^3 = \frac{\varphi_0\alpha_0^3}{\Phi_0}\;,
\end{equation}
to calculate $\varphi_0$ where necessary. 

For a inclusion made of a neo-Hookean material, the total dimensionless elastic free energy of the inclusion-matrix system can be given as
\begin{equation}\label{eq:PsiSpheroid}
    \begin{split}
        \bar{\Psi} =\;& \;\; \int_1^\infty \frac{\Phi}{\Phi_0}\Lambda^2\left[\;2\;(\lambda^2-\bar{\lambda}^2)\left(1-\frac{1}{\lambda^2\bar{\lambda}^4}\right) +\left(2+\frac{\Phi_\infty^2}{\Phi^2}\right)\left(2\bar{\lambda}^2+\frac{1}{\bar{\lambda}^4}-3\right)\right.\\
        &\underbrace{\left.+\;\frac{(\Phi_\infty^2-1)}{\Phi^2}\left(\frac{\Phi_\infty^2}{\lambda^2\bar{\lambda}^4}-1\right)\frac{(\lambda-\bar{\lambda})^2}{\gamma^3}\left(2\gamma^2+3\gamma+\frac{3}{2}\frac{\Phi_\infty^2}{\Phi^2}\sqrt{-\gamma}\ln\left(\frac{1+\sqrt{-\gamma}}{1-\sqrt{-\gamma}}\right)\right)\right]\mathrm{d}\Lambda}_{\displaystyle\bar\Psi_M}\;\\
        &+ (\lambda^*)^3\left[2\left(\frac{\Phi_0}{\varphi_0}\right)^{\frac{2}{3}}+\left(\frac{\varphi_0}{\Phi_0}\right)^{\frac{4}{3}}-3\right],
    \end{split}
\end{equation}
where $\bar\Psi_M$ is the dimensionless elastic free energy of only the matrix. This expression is agnostic to whether the spheroidal inclusion is prolate or oblate (while $\sqrt{-\gamma}$ might be imaginary, $\bar{\Psi}$ will always be real). 
\footnote{Remembering that for our choice of $\hat{\Phi}(\Lambda)$, from (\ref{eq:gammaSub}) $\gamma = (\Phi_\infty^2/\Phi^2-1)$, in the case of the oblate spheroid $\gamma>0$ and the logarithmic term in (\ref{eq:PsiSpheroid}) becomes an inverse tangent term. In the case of the prolate spheroid $\gamma<0$ and the logarithmic term in (\ref{eq:PsiSpheroid}) becomes a inverse hyperbolic tangent term.}

The parameter $\Phi_\infty$ can now be determined for a given  $\lambda^*$ and $\Phi_0$ by satisfying (\ref{eq:minimizationConditions}) via the single condition
\begin{equation}
     \frac{\partial}{\partial\Phi_{\infty}}\bar{\Psi}(\lambda^*,\Phi_{0},\Phi_{\infty})=0\;.
\end{equation}
Finally, the pressure $p_I$ inside the inclusion can be determined by
\begin{equation}\label{eq:qSpheroid}
    \frac{p_I}{\mu} = \frac{1}{2}\frac{\partial\bar\Psi_M}{\partial J^*_I} + 2\;(\lambda^*)^2\left[\left(\frac{\Phi_0}{\varphi_0}\right)^{\frac{1}{3}}-\left(\frac{\varphi_0}{\Phi_0}\right)^{\frac{5}{3}}\right]\left(\;\frac{\partial\alpha_0}{\partial J^*_I}-\frac{1}{3}\frac{\alpha_0}{J^*_I}\right)\;.
\end{equation}
Thus, we have fully determined the stress state inside the inclusion. 
\end{subequations}
Once $\Phi_\infty$ is determined, the form of the deformation field is entirely analytic. However, because the authors were unable to analytically evaluate the integral necessary to calculate $\bar\Psi_M$, $\Phi_\infty$ was determined numerically. Maybe the reader will have the luck or skill necessary to attain fully analytical results. 

In order to validate the results attained for spheroids, we performed axisymmetric finite element analysis (FEA) of an isotropically growing inclusion using the open source software FEniCS, and the framework found in \cite{senthilnathan2024understanding} (see Section 5.5.1, Appendix C and Appendix E therein) with the following, nearly incompressible neo-Hookean elastic free energy function
\begin{equation}
    \psi^*(I_1,J^e) = \frac{\mu}{2}(I_1-3-2\ln(J^e))+\frac{\kappa}{2}(\ln(J^e))^2, 
\end{equation}
where we set $\kappa/\mu = 1000$ to approximate incompressible behavior. The FEA results are compared with the Eshelby's linear solution \cite{eshelby1957,eshelby1959}, a nonlinear solution using the confocal ellipses $\hat\Phi^C$ (as used in \cite{PorousEllipsoids,GologanuOblate,GologanuProlate}), and the generalized solution formulated in this work using $\hat\Phi^G$.

\begin{figure}[t]
    \centering
    \includegraphics[width=0.98\textwidth]{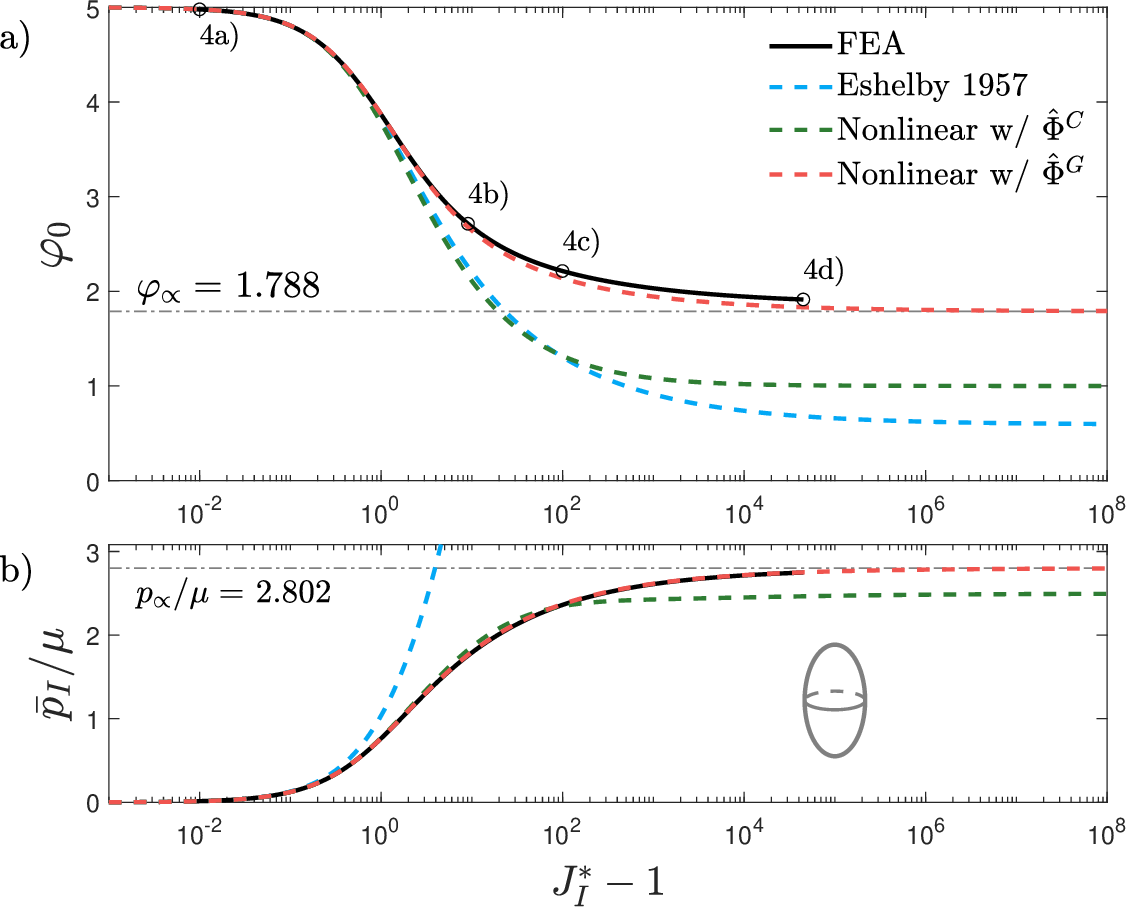}
    \caption{The Deformed aspect ratio of the inclusion $\varphi_0$ (a) and volume averaged dimensionless inclusion pressure $\bar{p}_I/\mu$ (b) vs. the volume growth ratio ($J^*_I-1$) of an incompressible and , neo-Hookean, isotropically growing prolate spheroidal inclusion-matrix system with undeformed aspect ratio $\Phi_0 = 5$. The dashed grey lines in (a) and (b) denote the isomorphic aspect ratio of the inclusion-matrix system $\varphi_\propto  = 1.788$ and the dimensionless isomorphic inclusion pressure of the inclusion-matrix system $p_\propto/\mu= 2.802$ respectively (as derived in Section \ref{sec:4.3}). The numerical labels in (a) denote the volume growth ratios of the deformation magnitude plots in Fig. \ref{fig:5_fullfield}.}
    \label{fig:5_2D}
\end{figure}
\begin{figure}[b]
    \centering
    \includegraphics[width=0.83\textwidth]{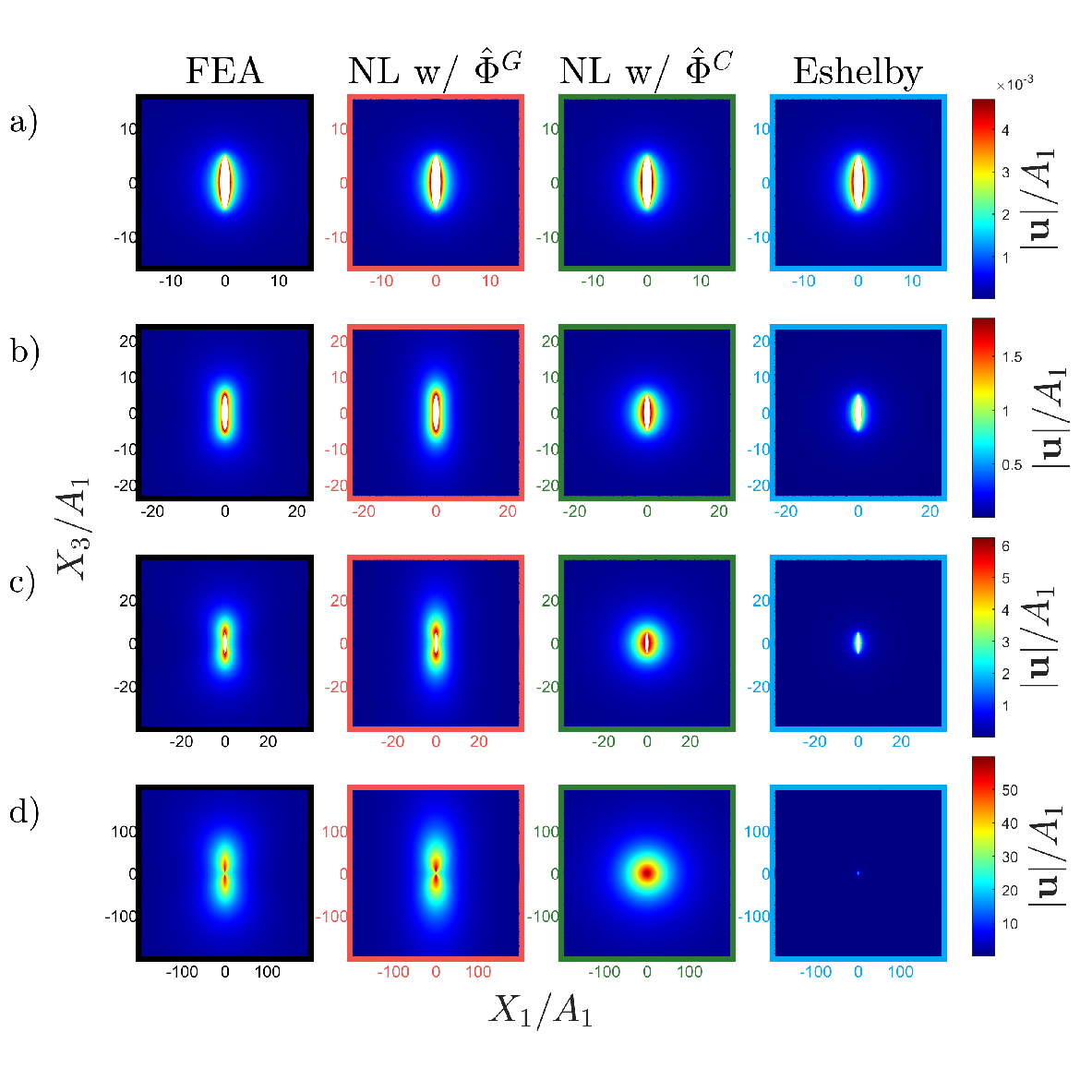}
    \caption{Full field plots of the normalized displacement magnitude $|\mathbf{u}|/A_1$ in the normalized reference coordinates $\X/A_1$ of an incompressible, neo-Hookean, isotropically growing prolate spheroidal inclusion-matrix system with undeformed aspect ratio $\Phi_0 = 5$. The volume growth ratios in each set of subplots are are (a) $J^*_I = 1.01$, (b) $J^*_I = 10$, (c) $J^*_I = 100$, and (d) $J^*_I = 26424$. Note that the colorbar scale and zoom level vary between each set of plots.}
    \label{fig:5_fullfield}
\end{figure}
\begin{figure}[t]
    \centering
    \includegraphics[width=0.98\textwidth]{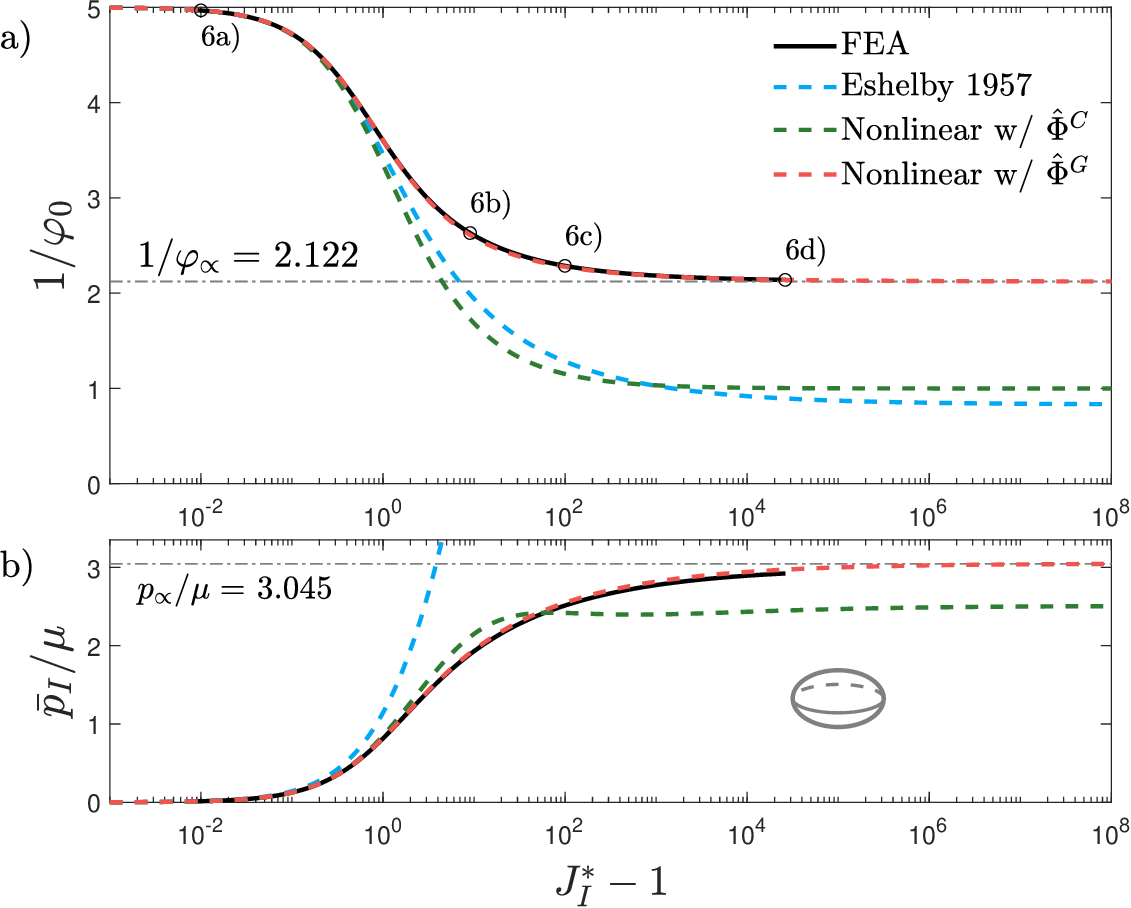}
    \caption{The deformed aspect ratio of the inclusion $\varphi_0$ (a) and volume averaged dimensionless inclusion pressure  $\bar{p}_I/\mu$ (b) vs. the volume growth ratio ($J^*_I-1$) of an incompressible and , neo-Hookean, isotropically growing prolate spheroidal inclusion-matrix system with undeformed aspect ratio $\Phi_0 = 1/5$. The dashed grey lines in (a) and (b) denote the isomorphic aspect ratio of the inclusion-matrix system $\varphi_\propto  = 0.471$ and the dimensionless isomorphic inclusion pressure of the inclusion-matrix system $p_\propto/\mu= 3.045$ respectively (as derived in Section \ref{sec:4.3}). The numerical labels in (a) denote the volume growth ratios of the deformation magnitude plots in Fig. \ref{fig:0.2_fullfield}.}
    \label{fig:0.2_2D}
\end{figure}
\begin{figure}[b]
    \centering
    \includegraphics[width=0.83\textwidth]{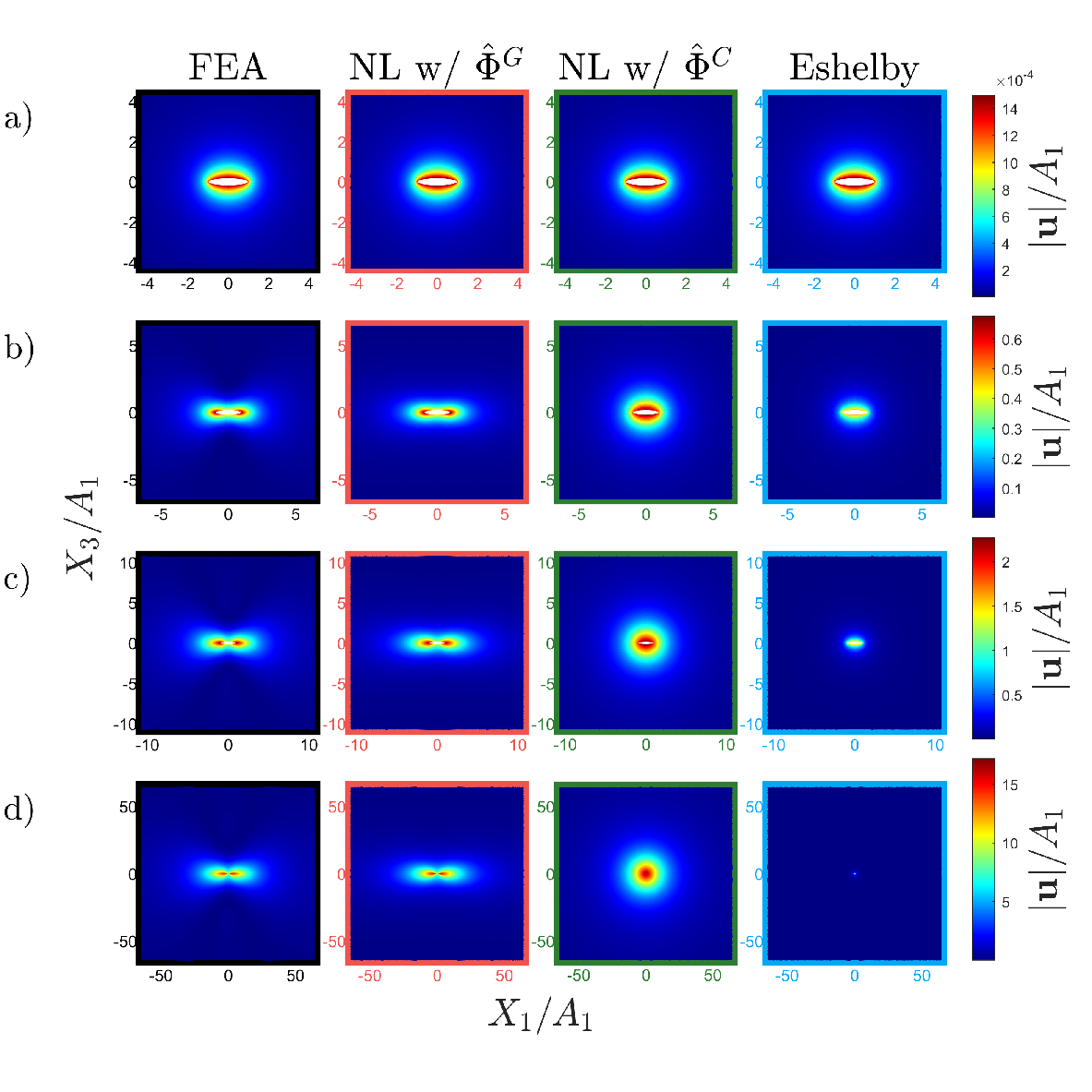}
    \caption{Full field plots of the normalized displacement magnitude $|\mathbf{u}|/A_1$ in the normalized reference coordinates $\X/A_1$ of an incompressible, neo-Hookean, isotropically growing oblate spheroidal inclusion-matrix system with undeformed aspect ratio $\Phi_0 = 1/5$. The volume growth ratios in each set of subplots are are (a) $J^*_I = 1.01$, (b) $J^*_I = 10$, (c) $J^*_I = 100$, and (d) $J^*_I = 45289$. Note that the colorbar scale and zoom level vary between each set of plots.}
    \label{fig:0.2_fullfield}
\end{figure}

The results for a prolate spheroid with $\Phi_0=5$, are shown in Figs. \ref{fig:5_2D} and \ref{fig:5_fullfield}, and the results for an oblate spheroid with $\Phi_0=1/5$ are shown in Figs. \ref{fig:0.2_2D} and \ref{fig:0.2_fullfield}. The results for spheroids of aspect ratios $\Phi_0=10$, $2$, $1/2$, and $1/10$ are shown in Appendix C. Figs. \ref{fig:5_2D} and \ref{fig:0.2_2D} a) present plots of the deformed aspect ratio of the inclusion $\varphi_0$ as the inclusion grows (increasing $J^*_I$). Figures \ref{fig:5_2D} and \ref{fig:0.2_2D} b) present plots of the volume averaged pressure inside the inclusion $\bar{p}_I$ nondimensionalized by the ground state shear modulus of the material $\mu$ as the inclusion grows.\footnote{We use the volume averaged pressure inside the inclusion, because in the simulations the deformation was \emph{not} uniform inside the inclusion, and therefore although $p_I = \bar{p}_I$ in each of the analytical models, in order to compare directly to the FEA model it was necessary to take the average pressure.} In each plot, the solution presented in this work demonstrates striking accuracy and outperforms both the linear and confocal model both qualitatively and quantitatively.

Most importantly, note that both the FEA and the $\hat\Phi^G$ model tend towards some non-spherical shape with an aspect ratio between $\Phi_0$ and $1$, and some pressure greater than the cavitation limit\cite{gentCav} of a neo-Hookean material of  $5\mu/2$. This behavior is not captured by any other existing model for the system. The Eshelby solution, incorrectly inverts from a prolate to oblate (or vice versa) deformed inclusion shape and the inclusion pressure increases boundlessly with $J^*_I$.  The confocal solution, on the other hand, incorrectly approaches a spherical shape and the spherical cavitation pressure exactly, even for non spherical inclusions. 
Note that another useful generalization of the confocal set, which was introduced in\cite{PhotoCuredEshelby} has the same limitations.

Furthermore, Figs. \ref{fig:5_fullfield} and \ref{fig:0.2_fullfield}, which plot the magnitude of the normalized displacement $|\mathbf{u}|/A_1 = |\x-\X|/A_1$ in the normalized reference coordinates $\X/A_1$ at several orders of magnitude of $J^*_I$, demonstrate the ability of our theory to capture the behavior of the FEA simulations in a way that the linear solution and confocal solution do not. For small $J^*_I$, the models all agree. Energy minimization shows that in this linear limit, $\Phi_\infty\to1$, and thus the three fields are equivalent as they all must obey incompressibility and satisfy energy minimization. However, as $J^*_I$ increases, the linear solution has a much smaller field of influence on the body, as it can only scale with the transformation strain of the inclusion. The confocal solution on the other hand, has a similarly sized field of influence of the deformation to that of the FEA but becomes spherical as the inclusion grows larger and larger. The shape of the field of influence of the deformation of FEA and solution presented in this work begin with a multi-lobed shape, and approach a shape with aspect ratio $\Phi_\infty\neq1$ far from the inclusion. Similar observations were noted in \cite{PhotoCuredEshelby}. This means that $\Phi_\infty\neq1$ dictates the ``shape'' of the deformation in the far field. Regardless of the value of $\Phi_\infty$ the deformation will vanish as $\Lambda\to\infty$, but the way in which it does so is dictated by this parameter. The reasons for this will be elucidated in the following section.

\section{The Isomorphic Limit}\label{sec:4.3}
In the following section we investigate the behavior of an isotropically growing inclusion as its volume tends towards infinity, implying $J^*_I\to\infty$ and by extension $\dV\to\infty$. We can equivalently imagine this as the limit where the inclusions initial volume goes to zero, i.e. $V^0/V^*\to0$. We find that for a neo-Hookean material an arbitrary ellipsoidal inclusion asymptotically approaches a non-spherical shape. Because in this limit the inclusion approaches this constant shape, we call this limit the isomorphic limit and the associated aspect ratios and maximum pressure (analogous to the well established cavitation pressure)  the isomorphic aspect ratios and isomorphic pressure, respectively. 

\subsection{The Isomorphic Aspect Ratios}
In this limit, because we assume $\Phi_2$ and $\Phi_3$ are finite, $\Phi_2\Phi_3\Lambda^2$ is negligible unless $O(\Lambda^2)\geq O(\Delta)$. As $\Lambda\to\infty$, $\Phi_J\to\Phi_{\infty J}$, thus, as $\Delta\to\infty$, (\ref{eq:alphaMatGen}) becomes
\begin{equation}\label{eq:alphaMatGenIso}
    \lim_{\Delta\to\infty}\ln(\alpha)=\lim_{\Delta\to\infty}\int{\frac{\Phi_{\infty 2}\Phi_{\infty 3}\Lambda^2}{\dV+\Phi_{\infty 2}\Phi_{\infty 3}\Lambda^3}\;\mathrm{d}\Lambda}\;, \;\forall\;\X\in\mathcal{D}^0_M\;.
\end{equation}
Notice that this implies that the deformation field inside the matrix no longer depends on the shape of the region which the inclusion initially occupied. Thus, in this limit our kinematic assumptions imply the deformation field $\displaystyle\lim_{\Delta\to\infty}\Chi(\X)$ can be described via the following functions for the undeformed aspect ratios
\begin{subequations}\label{eq:fullsetSpheroidIso}
    \begin{equation}\label{eq:PhiMatSpheroidIso}
        \lim_{\Delta\to\infty}\hat\Phi_J(\Lambda)=
        \begin{dcases}
            \Phi_{0J}, & 0<\Lambda< 1 \\
            \varphi_{\propto J}, & \Lambda>1\;,
        \end{dcases}
    \end{equation}
deformed aspect ratios
    \begin{equation}\label{eq:phiMatSpheroidIso}
        \lim_{\Delta\to\infty}\hat\varphi_J(\Lambda)=\varphi_{\propto J}\;,\forall\; \Lambda\;,
    \end{equation}
and deformed radial parameter
    \begin{equation}\label{eq:alphaMatSpheroidIso}
        \lim_{\Delta\to\infty}\hat\alpha(\Lambda)=
        \begin{dcases}
            \lambda^*\left(\frac{\Phi_{02}\Phi_{03}}{\varphi_{\propto2}\varphi_{\propto3}}\right)^{\frac{1}{3}}\Lambda, & 0<\Lambda< 1 \\
            \left(\Lambda^3 + \frac{\Delta}{\varphi_{\propto2}\varphi_{\propto3}}\right)^{\frac{1}{3}}, & \Lambda>1\;.
        \end{dcases}
    \end{equation}
    \end{subequations}
Because the deformation field can be described by a constant deformed aspect ratios everywhere we call this large growth limit the \emph{isomorphic limit} and the corresponding aspect ratios $\varphi_{\propto J}$ as the isomorphic aspect ratios. The value of each $\varphi_{\propto J}$ can now be determined via energy minimization. 

Here, we also find an explanation for why the confocal solution becomes spherical as $J^*_I\to\infty$: it is constrained to do so by incompressibility since $\varphi_\propto = \Phi_\infty=1$. Therefore the solution cannot capture the non-spherical asymptotic behavior seen in the simulations.


\subsection{Energy Minimization and the Isomorphic Pressure}
Through the calculations in Appendix B it can be shown that for an arbitrary ellipsoidal inclusion, in the isomorphic limit, the total dimensionless free energy in the inclusion-matrix system is 
\begin{equation}\label{eq:PsiIso}
    \begin{split}
        \tilde{\Psi}=\lim_{\dV\to\infty}\frac{2\Psi}{\mu V^*}=&\underbrace{\frac{1}{5}\left[\frac{(\varphi_{\propto2}^2-1)^2}{\varphi_{\propto2}^2}+\frac{(\varphi_{\propto3}^2-1)^2}{\varphi_{\propto3}^2}+\left(\frac{\varphi_{\propto2}}{\varphi_{\propto3}}-\frac{\varphi_{\propto3}}{\varphi_{\propto2}}\right)^2\right]+5}_{\displaystyle\tilde\Psi_M}\\
        & +\left[\left(\frac{\Phi_{02}\Phi_{03}}{\varphi_{\propto2}\varphi_{\propto3}}\right)^{\frac{2}{3}}+\left(\frac{\varphi_{\propto2}^2\Phi_{03}}{\Phi_{02}^2\varphi_{\propto3}}\right)^{\frac{2}{3}}+\left(\frac{\Phi_{02}\varphi_{\propto3}^2}{\varphi_{\propto2}\Phi_{03}^2}\right)^{\frac{2}{3}}-3\right]\;,
    \end{split}
\end{equation}
where $\tilde{\Psi}_M$ is the total dimensionless free energy inside the matrix. Notice that by normalizing by the grown volume the formulation for an arbitrary ellipsoidal inclusion becomes \emph{scale invariant} and no longer depends on $J^*_I$, $\lambda^*$, $\dV$ or any other related quantity. In order to identify the values of $\varphi_{\propto 2}$ and $\varphi_{\propto 3}$ which minimize the elastic free energy, we find the fixed point where $\partial\tilde\Psi/\partial \varphi_{\propto 2}=\partial\tilde\Psi/\partial \varphi_{\propto 3}=0$. In Appendix C it is further shown that as long as $\Phi_{02}\neq0$ or $\Phi_{03}\neq0$ there exists a non-spherical shape which minimizes the total free energy of the system. 

Any exact solution must have lower free energy than our approximate system, and since $\varphi_{\propto 2}=\varphi_{\propto 3}=1$ \emph{is not} the minimum energy solution for even our approximate form of deformation, the true solution as $J^*_I\to\infty$ must be different from that of isotropic spherical expansion. 

As the inclusion grows, in a neo-Hookean material, it will also reach a maximum pressure (analagous to a cavitation pressure). We call this maximum inclusion pressure $\displaystyle p_\propto = \lim_{\Delta\to\infty}p_I$ the isomorphic pressure and for an arbitrary ellipsoidal inclusion it can be found as
\begin{equation}\label{eq:pIso}
    \frac{p_\propto}{\mu} = \frac{\tilde\Psi_M}{2} = \frac{5}{2} + \frac{1}{10}\left[\frac{(\varphi_{\propto2}^2-1)^2}{\varphi_{\propto2}^2}+\frac{(\varphi_{\propto3}^2-1)^2}{\varphi_{\propto3}^2}+\left(\frac{\varphi_{\propto2}}{\varphi_{\propto3}}-\frac{\varphi_{\propto3}}{\varphi_{\propto2}}\right)^2\right].
\end{equation}
In the case of a spherical inclusion ($\varphi_{\propto2} = \varphi_{\propto3} = 1$) the isomorphic pressure reduces to the cavitation pressure $5\mu/2$. And in any other case, the isomorphic pressure is higher than the cavitation pressure. This means that a growing ellipsoidal inclusion must be able to sustain a higher pressure than the cavitation pressure to be able to grow indefinitely large, even in a neo-Hookean material. 

\subsection{Isomorphic Limit of a Spheroidal Inclusion}

\begin{figure}[t]
    \centering
    \includegraphics[width=0.9\textwidth]{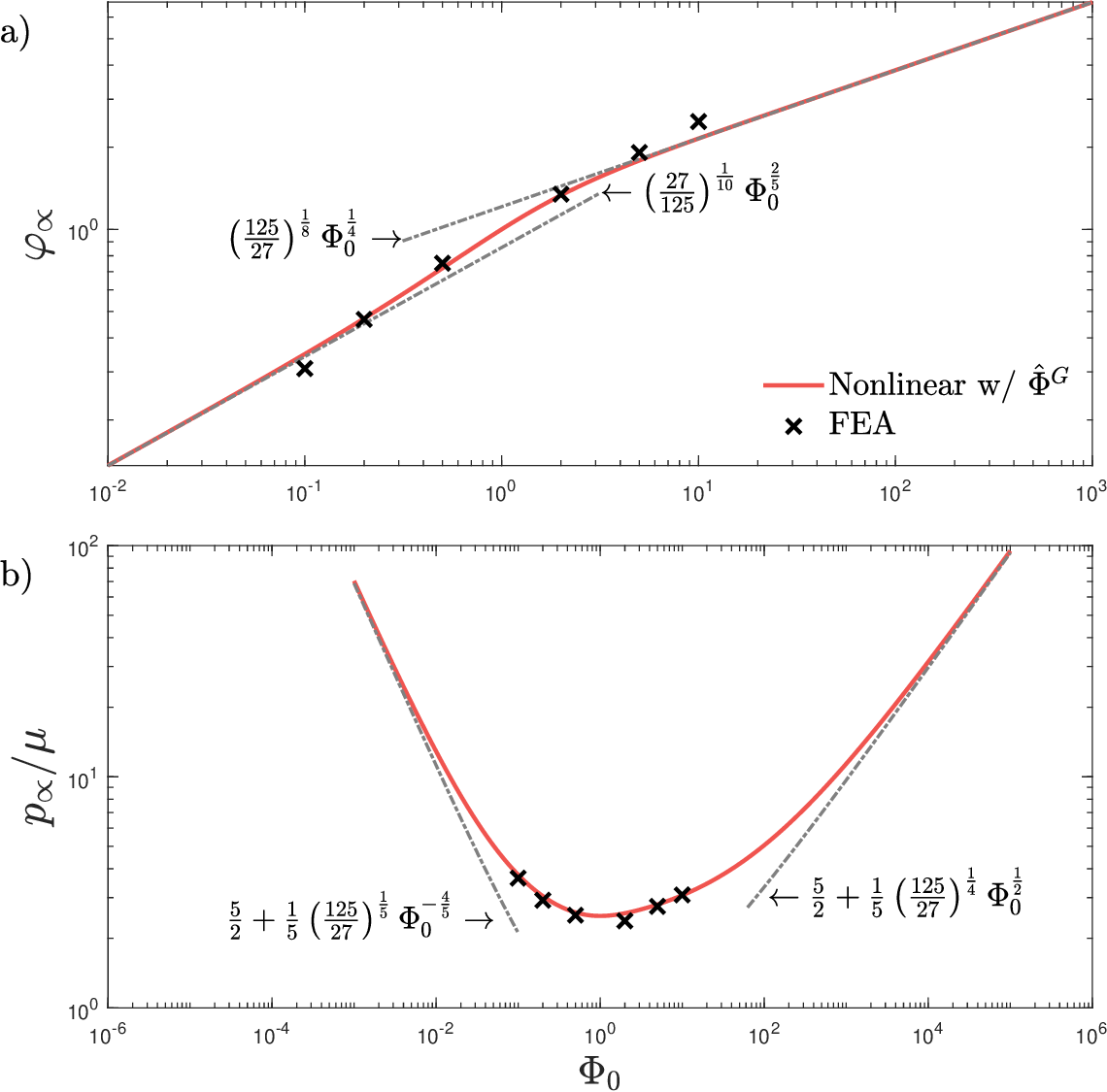}
    \caption{(a) Isormorphic aspect ratio $\varphi_\propto$ and (b) non-dimensional isomorphic inclusion pressure $p_\propto$ of a an incompressible, neo-Hookean, spheroidal inclusion-matrix system vs undeformed aspect ratio $\Phi_0$.}
    \label{fig:Isomorphic}
\end{figure}

It can trivially be seen that for a spheroidal inclusion, described by only one isomorphic aspect ratio $\varphi_\propto$, (\ref{eq:PsiIso}) reduces to
\begin{equation}\label{eq:PsiSpheroidIso}
    \tilde{\Psi}=\lim_{\dV\to\infty}\frac{2\Psi}{\mu V^*}=\underbrace{\frac{2}{5}\frac{(\varphi_{\propto}^2-1)^2}{\varphi_{\propto}^2}+5}_{\displaystyle\tilde\Psi_M}\;\;+\;\;2\left(\frac{\Phi_{0}}{\varphi_{\propto}}\right)^{\frac{2}{3}}+\left(\frac{\varphi_{\propto}}{\Phi_{0}}\right)^{\frac{4}{3}}-3\;,
\end{equation}
and (\ref{eq:pIso}) reduces to
\begin{equation}\label{eq:pIsoSpheroid}
    \frac{p_\propto}{\mu} = \frac{\tilde\Psi_M}{2} = \frac{5}{2} + \frac{1}{5}\frac{(\varphi_{\propto}^2-1)^2}{\varphi_{\propto}^2}\;.
\end{equation}
By setting $\partial\tilde\Psi/\partial \varphi_{\propto} = 0$ we find that that $\varphi_\propto$ can be found as any real positive root of the polynomial
\begin{equation}\label{eq:isomorphicEquation}
    P(\varphi_\propto)=27\Phi_0^4\left(\varphi_\propto^4-1\right)^3+125\varphi_\propto^4(\varphi_\propto^2-\Phi_0^2)^3\;.
\end{equation} 
From (\ref{eq:isomorphicEquation}) it can be seen by inspection that for the spherical case ($\Phi_0=1$)  that $\varphi_\propto$=1 is a real positive root of $P$, and that for any other positive value of $\varphi_\propto$, $P>0$ or $P<0$. Thus, for a spherical inclusion, we find that the isomorphic limit is also always spherical. 

From (\ref{eq:isomorphicEquation}) it can be seen by inspection that for oblate spheroids ($\Phi_0<1$), if $\varphi_\propto\leq\Phi_0$ then $P < 0$ and if $\varphi_\propto\geq1$ then $P > 0$. Because $P$ is a continuous function of $\varphi_\propto$ this means that $P$ must have at least one positive real root in the range $\varphi_\propto \in (\Phi_0, 1)$. Additionally, the first derivative of $P$ is given as
\begin{equation}\label{eq:isomorphicDerivative}
    P'(\varphi_\propto)=324\Phi_0^{4} \varphi_\propto^{3} \left(\varphi_\propto^{4} - 1\right)^{2} + 500\varphi_\propto^{3} \left(\varphi_\propto^{2} - \Phi_0^{2}\right)^{3} + 750\varphi_\propto^{5} \left(\varphi_\propto^{2} - \Phi_0^{2}\right)^2.
\end{equation}
By inspection it is clear that for an oblate spheroid with $\varphi_\propto \in (\Phi_0, 1)$, $P$ is monotonically increasing. Thus, for an oblate spheroid, we find that the isomorphic limit is also oblate with an aspect ratio between $\Phi_0$ and $1$.

Finally, from (\ref{eq:isomorphicEquation}) it can be seen by in inspection that for prolate spheroids ($\Phi_0>1$), if $\varphi_\propto\geq\Phi_0$ then $P > 0$ and if $\varphi_\propto\leq1$ then $P < 0$. Because $P$ is a continuous function of $\varphi_\propto$ this means that $P$ must have at least one positive real root in the range $\varphi_\propto\in(1,\Phi_0)$. It can further be proven via Sturm's theorem that there is only one real root within this range. Thus, for a prolate spheroid, we find that the isomorphic limit is also prolate with an aspect ratio between $1$ and $\Phi_0$. 

Here we make two important notes:
\begin{enumerate}
    \item This analysis no longer depends on the assumption of an isotropically growing inclusion, only that the growth is homogeneous inside the inclusion. The isomorphic results depend on only \emph{the final, grown, shape} of the inclusion.
    \item In the limit where $\Phi_0\to\infty$ the problem becomes similar to that of a growing cylindrical cavity. For a cylindrical cavity cavitation does not occur. We see similarly that there is no finite $p_\propto$ as $\Phi_0\to\infty$. This analysis gives some understanding of a connection between the two problems. 
\end{enumerate}

The analytical values of the isomorphic aspect ratios $\varphi_\propto$ and associated isomorphic pressures $p_\propto$ plotted on Figs. \ref{fig:5_2D} and \ref{fig:0.2_2D}, agree with the limits of the results in the previous section, and match quantitatively well with the FEA results. The analytical values of $\varphi_\propto$ and $p_\propto$ vs. initial aspect ratio $\Phi_0$ are shown in Fig. \ref{fig:Isomorphic}. The limiting values of these functions are also shown in Fig. \ref{fig:Isomorphic}.

\section{Conclusion}
The accurate approximate solution in this work can be expressed in an analytical form (once $\Phi_\infty$ is determined) and is able to capture the behavior of an isotropically growing neo-Hookean inclusion in a way that no previous theory has. Therefore, it may be useful in predicting the physics of incompatibility in synthetic and biological soft matter systems, such as the growth of bacterial biofilms and tumors, or the curing induced stress fields within light cured polymers. To attain this level of accuracy, however, the solution is limited to isotropic growth and relies on the ground state material properties of the matrix and inclusion being the same. In conjunction with homogenization methods, this and similarly derived solutions to related problems (fluid inclusions, inhomogeneities, stiffening inclusion/matrix materials, anisotropic growth, non-spheroidal geometries, external loads etc.) may serve as a basis to better understand the micromechanics of soft materials in the future. 

Furthermore, this work elucidates the asymptotic behavior of an ellipsoidal inclusion as it grows infinitely large. An analysis of this heretofore unexamined physics would be functionally impossible with only numerical simulations. This limit, associated with a higher asymptotic pressure than that of the commonly examined spherical cavitation limit, gives insight into how the shape of an inclusion affects the maximum stress it must be able to sustain in order to grow indefinitely. 

The authors would like to emphasize a principal insight garnered from the methods used in this work: \emph{there is no reason to believe that the simplest way to express the solution to a nonlinear problem is the same at all levels of deformation.} Expressing the solution presented in this work in a standard spheroidal coordinate system (convenient in the linear limit) would completely obscure the simplicity of the solution made possible by choosing a more generalized set of scaling spheroids. As the field looks towards understanding related problems on the incompatibility of soft systems, we reiterate that a carefully chosen arbitrary coordinate system may result in the discovery of more elegant and interpretable solutions.

\begin{acks}
J. B. would like to acknowledge the National Science Foundation Graduate Research Fellowship Program as his primary funding source. T. C. would like to acknowledge the support from the National Science Foundation under award number CMMI 1942016. C. S. would like to acknowledge the Lillian Gilbreth Postdoctoral Fellowship offered by Purdue University. 
\end{acks}

\bibliographystyle{SageV}
\bibliography{Main.bib}

\newcommand{\noopsort}[1]{} \newcommand{\printfirst}[2]{#1} \newcommand{\singleletter}[1]{#1} \newcommand{\switchargs}[2]{#2#1}
\begin{thebibliography}{10}
\providecommand{\url}[1]{\texttt{#1}}
\providecommand{\urlprefix}{URL }
\expandafter\ifx\csname urlstyle\endcsname\relax
  \providecommand{\doi}[1]{DOI:\discretionary{}{}{}#1}\else
  \providecommand{\doi}{DOI:\discretionary{}{}{}\begingroup \urlstyle{rm}\Url}\fi
\providecommand{\eprint}[2][]{\url{#2}}

\bibitem{residual}
Mukherjee S.
\newblock Influence of residual stress in failure of soft materials.
\newblock \emph{Mechanics Research Communications} 2022; 123: 103903.
\newblock \doi{https://doi.org/10.1016/j.mechrescom.2022.103903}.

\bibitem{cheng2009}
Cheng G, Tse J, Jain RK et~al.
\newblock Micro-environmental mechanical stress controls tumor spheroid size and morphology by suppressing proliferation and inducing apoptosis in cancer cells.
\newblock \emph{PLoS ONE} 2009; 4(2).
\newblock \doi{10.1371/journal.pone.0004632}.

\bibitem{mills2014}
Mills KL, Kemkemer R, Rudraraju S et~al.
\newblock Elastic free energy drives the shape of prevascular solid tumors.
\newblock \emph{PLoS ONE} 2014; 9(7).
\newblock \doi{10.1371/journal.pone.0103245}.

\bibitem{zhang2021}
Zhang Q, Li J, Nijjer J et~al.
\newblock Morphogenesis and cell ordering in confined bacterial biofilms.
\newblock \emph{Proceedings of the National Academy of Sciences} 2021; 118(31): e2107107118.
\newblock \doi{10.1073/pnas.2107107118}.

\bibitem{plants}
Königsberger M, Lukacevic M and Füssl J.
\newblock Multiscale micromechanics modeling of plant fibers: Upscaling of stiffness and elastic limits from cellulose nanofibrils to technical fibers.
\newblock \emph{Materials and Structures} 2023; 56(1).
\newblock \doi{10.1617/s11527-022-02097-2}.

\bibitem{skin}
Yuan Z, Huang Q, Liang X et~al.
\newblock {A Constitutive Model of Human Dermis Skin Incorporating Different Collagen Fiber Families}.
\newblock \emph{Journal of Applied Mechanics} 2022; 89(4): 041007.
\newblock \doi{10.1115/1.4053360}.

\bibitem{holzapfel_gasser}
Holzapfel GA, Gasser TC and Ogden RW.
\newblock A new constitutive framework for arterial wall mechanics and a comparative study of material models.
\newblock \emph{Journal of Elasticity} 2000; 61(1/3): 1–48.
\newblock \doi{10.1023/a:1010835316564}.

\bibitem{PCPdefects1}
Liu T, Guessasma S, Zhu J et~al.
\newblock Microstructural defects induced by stereolithography and related compressive behaviour of polymers.
\newblock \emph{Journal of Materials Processing Technology} 2018; 251: 37--46.
\newblock \doi{https://doi.org/10.1016/j.jmatprotec.2017.08.014}.

\bibitem{PCPdefects2}
Hook AL, Scurr DJ, Burley JC et~al.
\newblock Analysis and prediction of defects in uv photo-initiated polymer microarrays.
\newblock \emph{J Mater Chem B} 2013; 1: 1035--1043.
\newblock \doi{10.1039/C2TB00379A}.

\bibitem{PhotoCuredEshelby}
Zhang Q, Shi Y and Gao C.
\newblock The determination of the curing induced, nonlinear elastic field of an inclusion in photo-cured materials.
\newblock \emph{International Journal of Solids and Structures} 2024; 302: 112978.
\newblock \doi{https://doi.org/10.1016/j.ijsolstr.2024.112978}.

\bibitem{eshelby1957}
Eshelby JD.
\newblock The determination of the elastic field of an ellipsoidal inclusion, and related problems.
\newblock \emph{Proceedings of the Royal Society of London Series A, Mathematical and Physical Sciences} 1957; 241(1226): 376--396.
\newblock \doi{https://doi.org/10.1098/rspa.1957.0133}.

\bibitem{eshelby1959}
Eshelby JD.
\newblock The elastic field outside an ellipsoidal inclusion.
\newblock \emph{Proceedings of the Royal Society of London Series A, Mathematical and Physical Sciences} 1959; 252(1271).
\newblock \doi{https://doi.org/10.1098/rspa.1959.0173}.

\bibitem{mura}
Mura T.
\newblock \emph{Micromechanics of Defects in Solids}.
\newblock 2nd ed. Springer Dordrecht, 1987.
\newblock \doi{https://doi.org/10.1007/978-94-009-3489-4}.

\bibitem{moritanakacomposites}
Benveniste Y.
\newblock A new approach to the application of mori-tanaka's theory in composite materials.
\newblock \emph{Mechanics of Materials} 1987; 6(2): 147--157.
\newblock \doi{https://doi.org/10.1016/0167-6636(87)90005-6}.

\bibitem{composites}
Jarali CS, Madhusudan M, Vidyashankar S et~al.
\newblock A new micromechanics approach to the application of eshelby's equivalent inclusion method in three phase composites with shape memory polymer matrix.
\newblock \emph{Composites Part B: Engineering} 2018; 152: 17--30.
\newblock \doi{https://doi.org/10.1016/j.compositesb.2018.06.028}.

\bibitem{metamaterials}
Willis J.
\newblock From statics of composites to acoustic metamaterials.
\newblock \emph{Philosophical Transactions of the Royal Society A: Mathematical, Physical and Engineering Sciences} 2019; 377: 20190099.
\newblock \doi{10.1098/rsta.2019.0099}.

\bibitem{piezo1997}
Kuo WS and Huang JH.
\newblock On the effective electroelastic properties of piezoelectric composites containing spatially oriented inclusions.
\newblock \emph{International Journal of Solids and Structures} 1997; 34(19): 2445--2461.
\newblock \doi{https://doi.org/10.1016/S0020-7683(96)00154-0}.

\bibitem{SharmaPiezo}
Rahmati AH, Liu L and Sharma P.
\newblock Homogenization of electrets with ellipsoidal microstructure and pathways for designing piezoelectricity in soft materials.
\newblock \emph{Mechanics of Materials} 2022; 173: 104420.
\newblock \doi{10.1016/j.mechmat.2022.104420}.

\bibitem{liping}
Yuan T and Liu L.
\newblock Polynomial inclusions: Definitions, applications, and open problems.
\newblock \emph{Journal of the Mechanics and Physics of Solids} 2023; 181: 105440.
\newblock \doi{https://doi.org/10.1016/j.jmps.2023.105440}.
\newblock \urlprefix\url{https://www.sciencedirect.com/science/article/pii/S0022509623002442}.

\bibitem{yavarianisotropic}
Yavari A.
\newblock On eshelby’s inclusion problem in nonlinear anisotropic elasticity.
\newblock \emph{Journal of Micromechanics and Molecular Physics} 2021; 06: 2150002.
\newblock \doi{10.1142/S2424913021500028}.

\bibitem{Golgoon_Yavari_2017}
Golgoon A and Yavari A.
\newblock Nonlinear elastic inclusions in anisotropic solids.
\newblock \emph{Journal of Elasticity} 2017; 130(2): 239–269.
\newblock \doi{10.1007/s10659-017-9639-0}.

\bibitem{yandgisotropic}
Yavari A and Goriely A.
\newblock Nonlinear elastic inclusions in isotropic solids.
\newblock \emph{Proceedings Mathematical, physical, and engineering sciences / the Royal Society} 2013; 469: 20130415.
\newblock \doi{10.1098/rspa.2013.0415}.

\bibitem{HarmonicVoid2024}
Wang X and Schiavone P.
\newblock An elliptical incompressible liquid inclusion in a compressible hyperelastic solid of harmonic type.
\newblock \emph{Journal of Elasticity} 2024; \doi{10.1007/s10659-024-10074-9}.

\bibitem{HarmonicStiff2024}
Kui M, Dai M and Gao CF.
\newblock Elliptic rigid inclusion in soft materials of harmonic type.
\newblock \emph{Journal of Applied Mechanics} 2024; 91: 071004.
\newblock \doi{10.1115/1.4065160}.

\bibitem{parks}
{Lambert Diani} J and Parks D.
\newblock Problem of an inclusion in an infinite body, approach in large deformation.
\newblock \emph{Mechanics of Materials} 2000; 32(1): 43--55.
\newblock \doi{https://doi.org/10.1016/S0167-6636(99)00015-0}.

\bibitem{phdthesisSemiInverse}
De~Pascalis R.
\newblock \emph{The Semi-Inverse Method in solid mechanics: Theoretical underpinnings and novel applications}.
\newblock PhD Thesis, Salento University, 2010.

\bibitem{HouAbeyaratne}
Hang-Sheng H and Abeyaratne R.
\newblock Cavitation in elastic and elastic-plastic solids.
\newblock \emph{Journal of the Mechanics and Physics of Solids} 1992; 40(3): 571--592.
\newblock \doi{https://doi.org/10.1016/0022-5096(92)80004-A}.

\bibitem{GologanuProlate}
Gologanu M, Leblond JB and Devaux J.
\newblock Approximate models for ductile metals containing non-spherical voids—case of axisymmetric prolate ellipsoidal cavities.
\newblock \emph{Journal of the Mechanics and Physics of Solids} 1993; 41(11): 1723--1754.
\newblock \doi{https://doi.org/10.1016/0022-5096(93)90029-F}.

\bibitem{GologanuOblate}
Gologanu M, Leblond JB and Devaux J.
\newblock {Approximate Models for Ductile Metals Containing Nonspherical Voids—Case of Axisymmetric Oblate Ellipsoidal Cavities}.
\newblock \emph{Journal of Engineering Materials and Technology} 1994; 116(3): 290--297.
\newblock \doi{10.1115/1.2904290}.

\bibitem{PorousEllipsoids}
Avaz R and Naghdabadi R.
\newblock Effective behavior of porous elastomers containing aligned spheroidal voids.
\newblock \emph{Acta Mechanica} 2013; 224.
\newblock \doi{10.1007/s00707-013-0853-y}.

\bibitem{li2022}
Li J, Kothari M, Chockalingam S et~al.
\newblock Nonlinear inclusion theory with application to the growth and morphogenesis of a confined body.
\newblock \emph{Journal of the Mechanics and Physics of Solids} 2022; 159: 104709.
\newblock \doi{https://doi.org/10.1016/j.jmps.2021.104709}.

\bibitem{Decomp}
Alhasadi M and Federico S.
\newblock Eshelby’s inclusion problem in large deformations.
\newblock \emph{Zeitschrift für angewandte Mathematik und Physik} 2021; 72.
\newblock \doi{10.1007/s00033-021-01594-8}.

\bibitem{myThesis}
Bonavia J.
\newblock \emph{A Semi-analytical Model for Nonlinear Elliptical Inclusions with Spherical Eigenstrains}.
\newblock Master's Thesis, Massachusetts Institute of Technology, 2023.

\bibitem{senthilnathan2024understanding}
Senthilnathan C.
\newblock \emph{Understanding the mechanics of growth: A large deformation theory for coupled swelling-growth and morphogenesis of soft biological systems}.
\newblock PhD Thesis, Massachusetts Institute of Technology, 2024.

\bibitem{gentCav}
Gent A.
\newblock Cavitation in rubber: a cautionary tale.
\newblock \emph{Rubber Chemistry and Technology} 1990; 63(3): 49--53.

\end{thebibliography}

  \renewcommand{\theequation}{A.\arabic{equation}}
  \setcounter{equation}{0}  
  \section*{Appendix A: Derivation of the Equations for a Spheroid}  
We have already established (\ref{eq:alphaMatSpheroid}), (\ref{eq:PhiMatSpheroid}), and (\ref{eq:JStarSpheroid}). Now all that is left to fully describe the field is to find the $\Phi_\infty$ which minimizes the total free energy in the system $\Psi$. In order to perform the necessary volume and area integrals we find it convenient to parametrize the undeformed and deformed coordinates as
\begin{subequations}
    \begin{equation}\label{eq:RefCoordsSpheroid}
       \X = A_1\Lambda\;\{\sin(v)\cos(u),\;\sin(v)\sin(u),\;\Phi\cos(v)\}\;,\;\mathrm{and}
    \end{equation}    
    \begin{equation}\label{eq:DefCoordsSpheroid}
        \x=A_1\alpha\;\{\sin(v)\cos(u),\;\sin(v)\sin(u),\;\varphi\cos(v)\}\;,
    \end{equation}  
\end{subequations}
respectively, which satisfy (\ref{eq:xofXSum}). In this way any point in the inclusion-matrix system $\X\in\mathcal{D}^0$ can be identified by its undeformed radial parameter $\Lambda$ and its two angular parameters $u\in[0,2\pi)$ and $v\in[0,\pi)$. By taking the Jacobian of (\ref{eq:RefCoordsSpheroid}) with respect to $\{\Lambda,u,v\}$ we find that the volume integral over $\mathcal{D}^0$ in the reference cooridinates can be expressed as
\begin{equation}\label{eq:volumeIntSpheroid}
    \int_{\mathcal{D}^0} \mathrm{dv}^0 = A_1^3\int_0^\infty\int_0^\pi\int_0^{2\pi} \left(\Phi\Lambda^2\sin(v)\Gamma\right)\mathrm{d}u\;\mathrm{d}v\;\mathrm{d}\Lambda\;,
\end{equation}
where for a spheroid in this parametrization $\Gamma = 1+\gamma \cos^2(v)$. By taking the Jacobian of (\ref{eq:DefCoordsSpheroid}) with respect to $\{u,v\}$ and evaluating at $\Lambda=1$ we find that the surface integral over $\partial\mathcal{D}_I$ in the deformed coordinates can be expressed as
\begin{equation}\label{eq:areaIntSpheroid}
    \int_{\partial\mathcal{D}_I} \mathrm{da} = \varphi_0\;a_1^2\int_0^\pi\int_0^{2\pi} \left(\sin(v)\sqrt{\sin^2(v)+\frac{1}{\varphi_0^2}\cos^2(v)}\right)\mathrm{d}u\;\mathrm{d}v\;.
\end{equation}

Via (\ref{eq:FjkSimpInc}) we can find the first invariant of $\B$ inside the inclusion
\begin{subequations}
    \begin{equation}\label{eq:I1IncSpheroid}
        I_1=2\left(\frac{\Phi_0}{\varphi_0}\right)^{\frac{2}{3}}+\left(\frac{\varphi_0}{\Phi_0}\right)^{\frac{4}{3}},\;\forall\;\X\in\mathcal{D}^0_I\;,
    \end{equation}
    and inserting (\ref{eq:gammaSub}) and (\ref{eq:RefCoordsSpheroid}) into (\ref{eq:FjkSimp}) and then using trigonometric identities to simplify, we can find the first invariant of $\B$ inside the matrix
    \begin{equation}\label{eq:I1MatSpheroid}
        \begin{split}
        I_1 \;= &\;\;\frac{(\Phi_\infty^2-1)}{\Phi^2}\left(\frac{\Phi_\infty^2}{\lambda^2\bar{\lambda}^4}-1\right)(\lambda-\bar{\lambda})^2\frac{\sin^2(v)\cos^2(v)}{\Gamma^{2}}\\
        & +(\lambda^2-\bar{\lambda}^2)\left(1-\frac{1}{\lambda^2\bar{\lambda}^4}\right)\frac{\sin^2(v)}{\Gamma} +\left(2\bar{\lambda}^2+\frac{1}{\bar{\lambda}^4}\right), \;\forall\;\X\in\mathcal{D}^0_M\;.
        \end{split} 
    \end{equation}
\end{subequations}
We can use the expressions in (\ref{eq:I1IncSpheroid}) and (\ref{eq:I1MatSpheroid}) to find the elastic free energy per unit volume in the ungrown reference via (\ref{eq:NHFE}) and (\ref{eq:FERV}) and find the expression for the total elastic free energy in the inclusion-matrix system via (\ref{eq:totalFreeEnergyArb}) and (\ref{eq:volumeIntSpheroid}) as
\begin{equation} \label{eq:totalFreeEnergySpheroid}
    \begin{split}
        \Psi =\;& \;\; \frac{\mu}{2}A_1^3\int_1^\infty\int_0^\pi\int_0^{2\pi} \Phi\Lambda^2\left[\frac{(\Phi_\infty^2-1)}{\Phi^2}\left(\frac{\Phi_\infty^2}{\lambda^2\bar{\lambda}^4}-1\right)(\lambda-\bar{\lambda})^2\frac{\sin^3(v)\cos^2(v)}{\Gamma}\right.\\
        &\left.+\;(\lambda^2-\bar{\lambda}^2)\left(1-\frac{1}{\lambda^2\bar{\lambda}^4}\right)\sin^3(v) +\left(2\bar{\lambda^2}+\frac{1}{\bar{\lambda}^4}-3\right)\sin(v)\Gamma\right]\mathrm{d}u\;\mathrm{d}v\;\mathrm{d}\Lambda\;\\
        &+ \frac{\mu}{2}(\lambda^*)^3\int_{\mathcal{D}^0_I}\left[2\left(\frac{\Phi_0}{\varphi_0}\right)^{\frac{2}{3}}+\left(\frac{\varphi_0}{\Phi_0}\right)^{\frac{4}{3}}-3\right]\;\mathrm{dv}^0\;.
    \end{split}
\end{equation}
Because $I_1$ is uniform inside the inclusion, the second integral in this expression is trivial and we can simply multiply the integrand by $V^0$. Once the second integral is evaluated, and the first integral is integrated over $u$ and $v$ we can apply (\ref{eq:gammaSub}) where convenient and substitute the result into (\ref{eq:psibarDef}), yielding (\ref{eq:PsiSpheroid}).

Next we calculate the pressure $p_I$ inside the inclusion, so that the stress field inside the inclusion is fully determined. In order to calculate $p_I$ we start by finding $\partial\x/\partial J^*_I$ and $\B^d\mathbf{n}$. $\partial\x/\partial J^*_I$ can be calculated trivially from (\ref{eq:JStarSpheroid}) and (\ref{eq:DefCoordsSpheroid}) at $\Lambda=1$, yielding
\begin{equation}\label{eq:dxdlstarSpheroid}
    \frac{\partial\x}{\partial J^*_I}=\;A_1\left\{\frac{\partial\alpha_0}{\partial J^*_I}\sin(v)\cos(u),\;\frac{\partial\alpha_0}{\partial J^*_I}\sin(v)\sin(u),\left(\frac{\alpha_0}{J^*_I}-2\frac{\partial\alpha_0}{\partial J^*_I}\;\right)\varphi_0\cos(v)\right\}.
\end{equation}
The outward unit normal $\mathbf{n}$ can be founding by taking the outward unit vector parallel to the cross product $(\partial{\x}/{\partial{u}}\times\partial{\x}/{\partial{v}})$ at $\Lambda=1$. Doing so yields
\begin{equation}
    \mathbf{n}=\frac{\left\{\sin(v)\cos(u),\;\sin(v)\sin(u),\;\frac{1}{\varphi_0}\cos(v)\right\}}{\sqrt{\sin^2(v)+\frac{1}{\varphi_0^2}\cos^2(v)}}.
\end{equation}
From (\ref{eq:FjkSimpInc}) we find that $\B^d\mathbf{n}$ thus equals
\begin{equation}\label{eq:BnSpheroid}
    \B^d\mathbf{n}=\frac{\left(\left(\frac{\Phi_0}{\varphi_0}\right)^{\frac{2}{3}}-\left(\frac{\varphi_0}{\Phi_0}\right)^{\frac{4}{3}}\right)\left\{\sin(v)\cos(u),\;\sin(v)\sin(u),\;-\frac{2}{\varphi_0}\cos(v)\right\}}{3\;\sqrt{\sin^2(v)+\frac{1}{\varphi_0^2}\cos^2(v)}}\;.
\end{equation}
Taking the dot product between (\ref{eq:dxdlstarSpheroid}) and (\ref{eq:BnSpheroid}) yields
\begin{equation}\label{eq:BndotSpheroid}
    \left(\B^d\mathbf{n} \cdot \frac{\partial\x}{\partial J^*_I}\right) =  A_1\left(\left(\frac{\Phi_0}{\varphi_0}\right)^{\frac{2}{3}}-\left(\frac{\varphi_0}{\Phi_0}\right)^{\frac{4}{3}}\right)\frac{\displaystyle\frac{\partial\alpha_0}{\partial J^*_I} \sin^2(v)-2\left(\frac{\alpha_0}{J^*_I}-2\frac{\partial\alpha_0}{\partial J^*_I}\right)\cos^2(v)}{3\;\sqrt{\sin^2(v)+\frac{1}{\varphi_0^2}\cos^2(v)}}\;.
\end{equation}
Integrating (\ref{eq:BndotSpheroid}) over $\partial\mathcal{D}_I$ via (\ref{eq:areaIntSpheroid}) yields
\begin{equation}\label{eq:BnIntegralSpheroid}
\int_{\partial\mathcal{D}_I}\left(\B^d\mathbf{n}\cdot\frac{\partial\mathbf{x}}{\partial J^*_I}\right)\mathrm{da} = \frac{8\pi}{3}\varphi_0 A_1a_1^2\left[\left(\frac{\Phi_0}{\varphi_0}\right)^{\frac{2}{3}}-\left(\frac{\varphi_0}{\Phi_0}\right)^{\frac{4}{3}}\right]\left(\;\frac{\partial\alpha_0}{\partial J^*_I}-\frac{1}{3}\frac{\alpha_0}{J^*_I}\right)
\end{equation}
Via (\ref{eq:VolumeInclusion}) and (\ref{eq:alphaMatSpheroid}) evaluated at $\Lambda=1$ we can simplify the prefactor and then insert (\ref{eq:BnIntegralSpheroid}) into (\ref{eq:genqEq}) to find
\begin{equation}
    p_I = \frac{1}{V^0}\left(\frac{\partial\Psi_M}{\partial J^*_I} + 2\mu V^0(\lambda^*)^2\left[\left(\frac{\Phi_0}{\varphi_0}\right)^{\frac{1}{3}}-\left(\frac{\varphi_0}{\Phi_0}\right)^{\frac{5}{3}}\right]\left(\;\frac{\partial\alpha_0}{\partial J^*_I}-\frac{1}{3}\frac{\alpha_0}{J^*_I}\right)\right).
\end{equation}
Dividing through by $\mu$ and $V^0$ yields (\ref{eq:qSpheroid}).
 \renewcommand{\theequation}{B.\arabic{equation}}
  \setcounter{equation}{0}  
  \section*{Appendix B: Derivation of the Equations for a General Ellipsoid in the Isomorphic Limit}  

We have already established (\ref{eq:fullsetSpheroidIso}). Now all that is left to fully describe the field is to find the $\Phi_\infty$ which minimizes the total free energy in the system $\Psi$. In order to perform the necessary volume and area integrals we find it convenient to parametrize the undeformed and deformed coordinates as
\begin{subequations}
    \begin{equation}\label{eq:RefCoordsIso}
       \X = A_1\Lambda\;\{\sin(v)\cos(u),\;\Phi_2\sin(v)\sin(u),\;\Phi_3\cos(v)\}\;,\;\mathrm{and}
    \end{equation}    
    \begin{equation}\label{eq:DefCoordsIso}
        \x=A_1\alpha\;\{\sin(v)\cos(u),\;\varphi_{\propto2}\sin(v)\sin(u),\;\varphi_{\propto3}\cos(v)\}\;,
    \end{equation}  
\end{subequations}
respectively which satisfy (\ref{eq:xofXSum}). In this way any point in the inclusion-matrix system $\X\in\mathcal{D}^0$ can be identified by its undeformed radial parameter $\Lambda$ and its two angular parameters $u\in[0,2\pi)$ and $v\in[0,\pi)$. By taking the Jacobian of (\ref{eq:RefCoordsIso}) with respect to $\{\Lambda,u,v\}$ we find that the volume integral over $\mathcal{D}^0$ in the reference cooridinates can be expressed as
\begin{equation}\label{eq:volumeIntIso}
    \int_{\mathcal{D}^0} \mathrm{dv}^0 = A_1^3\int_0^\infty\int_0^\pi\int_0^{2\pi} \left(\Phi_2\Phi_3\Lambda^2\sin(v)\right)\mathrm{d}u\;\mathrm{d}v\;\mathrm{d}\Lambda\;,
\end{equation}
 since (\ref{eq:PhiMatSpheroidIso}) and (\ref{eq:phiMatSpheroidIso}) imply that both $\gamma_J$ and $\beta_J$ vanish everywhere for $\Delta\to\infty$. This implies that $\Gamma\to1$ everywhere. By taking the Jacobian of (\ref{eq:DefCoordsIso}) with respect to $\{u,v\}$ and evaluating at $\Lambda=1$ we find that the surface integral over $\partial\mathcal{D}_I$ in the deformed coordinates can be expressed as
\begin{equation}\label{eq:areaIntIso}
    \int_{\partial\mathcal{D}_I} \mathrm{da} = \varphi_{\propto 2}\varphi_{\propto 3}\;a_1^2\int_0^\pi\int_0^{2\pi} \left(\sin(v)\textstyle\sqrt{\sin^2(v)\cos^2(u)+\frac{\sin^2(v)\sin^2(u)}{\varphi_{\propto2}^2}+\frac{\cos^2(v)}{\varphi_{\propto3}^2}}\right)\mathrm{d}u\;\mathrm{d}v\;.
\end{equation}

Through (\ref{eq:fullsetSpheroidIso}) we can the first invariant of $\B$ inside the inclusion
\begin{subequations}
    \begin{equation}\label{eq:I1IncIso}
        I_1=\left(\frac{\Phi_{02}\Phi_{03}}{\varphi_{\propto2}\varphi_{\propto3}}\right)^{\frac{2}{3}}+\left(\frac{\varphi_{\propto2}^2\Phi_{03}}{\Phi_{02}^2\varphi_{\propto3}}\right)^{\frac{2}{3}}+\left(\frac{\Phi_{02}\varphi_{\propto3}^2}{\varphi_{\propto2}\Phi_{03}^2}\right)^{\frac{2}{3}},\;\forall\;\X\in\mathcal{D}^0_I\;,
    \end{equation}
and inserting (\ref{eq:gammaSub}) and (\ref{eq:RefCoordsSpheroid}) into (\ref{eq:FjkSimp}) and then using trigonometric identities to simplify, we can find the first invariant of $\B$ inside the matrix
    \begin{equation}\label{eq:I1MatIso}
        \begin{split}
        I_1 \;= &\;\;\left[\frac{(\varphi_{\propto2}^2-1)^2}{\varphi_{\propto2}^3}X_1X_2+\frac{(\varphi_{\propto3}^2-1)^2}{\varphi_{\propto3}^3}X_1X_3+\left(\frac{\varphi_{\propto2}}{\varphi_{\propto3}}-\frac{\varphi_{\propto3}}{\varphi_{\propto2}}\right)^2\frac{X_2X_3}{\varphi_{\propto2}\varphi_{\propto3}}\right]\frac{(\lambda-\bar\lambda)^2}{A_1^2\Lambda^2}\\
        & +\left(2\bar\lambda^2+\frac{1}{\bar\lambda^4}\right), \;\forall\;\X\in\mathcal{D}^0_M\;.
        \end{split} 
    \end{equation} 
\end{subequations}

We can use the expressions in (\ref{eq:I1IncIso}) and (\ref{eq:I1MatIso}) to find the elastic free energy per unit volume in the ungrown reference via (\ref{eq:NHFE}) and (\ref{eq:FERV}) and find the expression for the total elastic free energy in the inclusion-matrix system via (\ref{eq:totalFreeEnergyArb}) and (\ref{eq:volumeIntIso}) as
\begin{equation}\label{eq:PsiMatIso}
        \begin{split}
            \Psi \;= &\;\;\frac{\mu}{2}\varphi_{\propto2}\varphi_{\propto3}A_1^3\int^\infty_1\left\{\frac{4\pi}{15}\left[\frac{(\varphi_{\propto2}^2-1)^2}{\varphi_{\propto2}^2}+\frac{(\varphi_{\propto3}^2-1)^2}{\varphi_{\propto3}^2}+\left(\frac{\varphi_{\propto2}}{\varphi_{\propto3}}-\frac{\varphi_{\propto3}}{\varphi_{\propto2}}\right)^2\right](\lambda-\bar\lambda)^2\right.\\ 
            & +4\pi\left(2\bar\lambda^2+\left.\frac{1}{\bar\lambda^4}-3\right)\right\}\Lambda^2\;\mathrm{d}\Lambda\\
            & + \frac{\mu}{2}(\lambda^*)^3 V^0\left[\left(\frac{\Phi_{02}\Phi_{03}}{\varphi_{\propto2}\varphi_{\propto3}}\right)^{\frac{2}{3}}+\left(\frac{\varphi_{\propto2}^2\Phi_{03}}{\Phi_{02}^2\varphi_{\propto3}}\right)^{\frac{2}{3}}+\left(\frac{\Phi_{02}\varphi_{\propto3}^2}{\varphi_{\propto2}\Phi_{03}^2}\right)^{\frac{2}{3}}-3\right],
        \end{split}
\end{equation} 
after integrating over $u$ and $v$. We can then integrate over $\Lambda$ and take the limit as $\Delta\to\infty$ yielding (\ref{eq:PsiIso}) remembering that in this limit $\Delta=\Phi_{02}\Phi_{03}J^*_I$.

Next we calculate the pressure $p_I$ inside the inclusion, so that the stress field inside the inclusion is fully determined. In order to calculate $p_I$ we start by finding $\partial\x/\partial J^*_I$ and $\B^d\mathbf{n}$. $\partial\x/\partial J^*_I$ can be calculated trivially from (\ref{eq:alphaMatSpheroidIso}) and (\ref{eq:DefCoordsIso}) at $\Lambda=1$, yielding
\begin{equation}\label{eq:dxdlstarIso}
    \frac{\partial\x}{\partial J^*_I}=\;\frac{A_1}{3(\lambda^*)^2}\left(\frac{\Phi_{02}\Phi_{03}}{\varphi_{\propto2}\varphi_{\propto3}}\right)^{\frac{1}{3}}\left\{\sin(v)\cos(u),\;\varphi_{\propto2}\sin(v)\sin(u),\;\varphi_{\propto3}\cos(v)\right\}.
\end{equation}
$\mathbf{n}$ can be founding by taking the outward unit vector parallel to the cross product $(\partial{\x}/{\partial{u}}\times\partial{\x}/{\partial{v}})$ at $\Lambda=1$. Doing so yields
\begin{equation}
    \mathbf{n}=\frac{\left\{\sin(v)\cos(u),\;\frac{1}{\varphi_{\propto2}}\sin(v)\sin(u),\;\frac{1}{\varphi_{\propto3}}\cos(v)\right\}}{\sqrt{\sin^2(v)\cos^2(u)+\frac{\sin^2(v)\sin^2(u)}{\varphi_{\propto2}^2}+\frac{\cos^2(v)}{\varphi_{\propto3}^2}}}.
\end{equation}
From (\ref{eq:FjkSimpInc}) we find that $\B^d\mathbf{n}$ thus equals
\begin{equation}\label{eq:BnIso}
    \begin{split}
        \B^d\mathbf{n}=& \left[\left(\frac{\Phi_{02}\Phi_{03}}{\varphi_{\propto2}\varphi_{\propto3}}\right)^{\frac{2}{3}}\left\{2\sin(v)\cos(u),\;-\frac{1}{\varphi_{\propto2}}\sin(v)\sin(u),\;-\frac{1}{\varphi_{\propto3}}\cos(v)\right\}\right.\\
        &+\left(\frac{\varphi_{\propto2}^2\Phi_{03}}{\Phi_{02}^2\varphi_{\propto3}}\right)^{\frac{2}{3}}\left\{-\sin(v)\cos(u),\;\frac{2}{\varphi_{\propto2}}\sin(v)\sin(u),\;-\frac{1}{\varphi_{\propto3}}\cos(v)\right\}\\
        &+\left.\left(\frac{\Phi_{02}\varphi_{\propto3}^2}{\varphi_{\propto2}\Phi_{03}^2}\right)^{\frac{2}{3}}\left\{-\sin(v)\cos(u),\;-\frac{1}{\varphi_{\propto2}}\sin(v)\sin(u),\;\frac{2}{\varphi_{\propto3}}\cos(v)\right\}\right]\\
        &*\textstyle\left(3\;\sqrt{\sin^2(v)\cos^2(u)+\frac{\sin^2(v)\sin^2(u)}{\varphi_{\propto2}^2}+\frac{\cos^2(v)}{\varphi_{\propto3}^2}}\right)^{-1}
    \end{split}
\end{equation}
Taking the dot product between (\ref{eq:dxdlstarIso}) and (\ref{eq:BnIso}) yields
\begin{equation}\label{eq:BndotIso}
      \begin{split}
        \left(\B^d\mathbf{n} \cdot \frac{\partial\x}{\partial J^*_I}\right) =& \left[\left(\frac{\Phi_{02}\Phi_{03}}{\varphi_{\propto2}\varphi_{\propto3}}\right)^{\frac{2}{3}}\left(2\sin^2(v)\cos^2(u)-\sin^2(v)\sin^2(u)-\cos^2(v)\right)\right.\\
        &+\left(\frac{\varphi_{\propto2}^2\Phi_{03}}{\Phi_{02}^2\varphi_{\propto3}}\right)^{\frac{2}{3}}\left(-\sin^2(v)\cos^2(u)+2\sin^2(v)\sin^2(u)-\cos^2(v)\right)\\
        &+\left.\left(\frac{\Phi_{02}\varphi_{\propto3}^2}{\varphi_{\propto2}\Phi_{03}^2}\right)^{\frac{2}{3}}\left(-\sin^2(v)\cos^2(u)-\sin^2(v)\sin^2(u)+2\cos^2(v)\right)\right]\\
        &*\textstyle A_1\left(\frac{\Phi_{02}\Phi_{03}}{\varphi_{\propto2}\varphi_{\propto3}}\right)^{\frac{1}{3}}\left(9(\lambda^*)^2\;\sqrt{\sin^2(v)\cos^2(u)+\frac{\sin^2(v)\sin^2(u)}{\varphi_{\propto2}^2}+\frac{\cos^2(v)}{\varphi_{\propto3}^2}}\right)^{-1}
    \end{split}
\end{equation}
Integrating (\ref{eq:BndotIso}) over $\partial\mathcal{D}_I$ via (\ref{eq:areaIntSpheroid}) yields
\begin{equation}\label{eq:BnIntegralIso}
\int_{\partial\mathcal{D}_I}\left(\B^d\mathbf{n}\cdot\frac{\partial\mathbf{x}}{\partial J^*_I}\right)\mathrm{da} = 0\;.
\end{equation}
Via (\ref{eq:VolumeInclusion}) and (\ref{eq:alphaMatGenIso}) evaluated at $\Lambda=1$ we can simplify the prefactor and then insert (\ref{eq:BnIntegralIso}) into (\ref{eq:genqEq}) to find
\begin{equation}
    p_\propto=\lim_{\Delta\to\infty} p_I = \frac{1}{V^0}\left(\frac{\partial\Psi_M}{\partial J^*_I}\right).
\end{equation}
Dividing through by $\mu$ and $V^0$ yields
\begin{equation}
    \frac{p_\propto}{\mu} = \frac{1}{2}\left(\frac{\partial\bar\Psi_M}{\partial J^*_I}\right)\;,
\end{equation}
and utilizing the fact that $\bar\Psi_M = J^*_I\tilde\Psi_M$ yields (\ref{eq:pIso}).
\newpage
\renewcommand{\theequation}{C.\arabic{equation}}
  \setcounter{equation}{0}  
  \renewcommand{\thefigure}{C\arabic{figure}}
  \setcounter{figure}{0}  
  \section*{Appendix C: Additional Figures}  

\begin{figure}[h]
    \centering
    \includegraphics[width=0.98\textwidth]{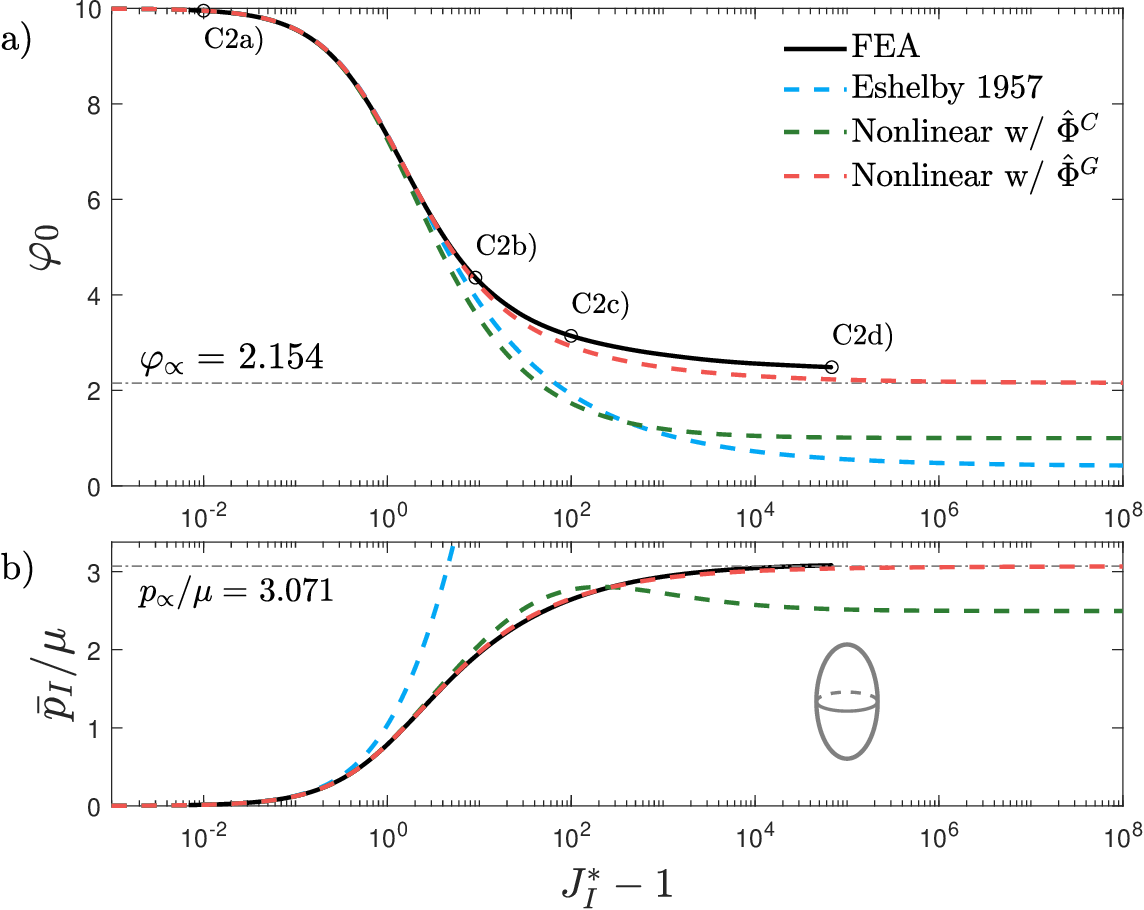}
    \caption{The deformed aspect ratio of the inclusion $\varphi_0$ (a) and volume averaged dimensionless inclusion pressure  $\bar{p}_I/\mu$ (b) vs. the volume growth ratio ($J^*_I-1$) of an incompressible and , neo-Hookean, isotropically growing prolate spheroidal inclusion-matrix system with undeformed aspect ratio $\Phi_0 = 10$. The dashed grey lines in (a) and (b) denote the isomorphic aspect ratio of the inclusion-matrix system $\varphi_\propto  = 2.154$ and the dimensionless isomorphic inclusion pressure of the inclusion-matrix system $p_\propto/\mu= 3.071$ respectively (as derived in Section \ref{sec:4.3}). The numerical labels in (a) denote the volume growth ratios of the deformation magnitude plots in Fig. \ref{fig:10_fullfield}.}
    \label{fig:10_2D}
\end{figure}
\begin{figure}[t]
    \centering
    \includegraphics[width=0.9\textwidth]{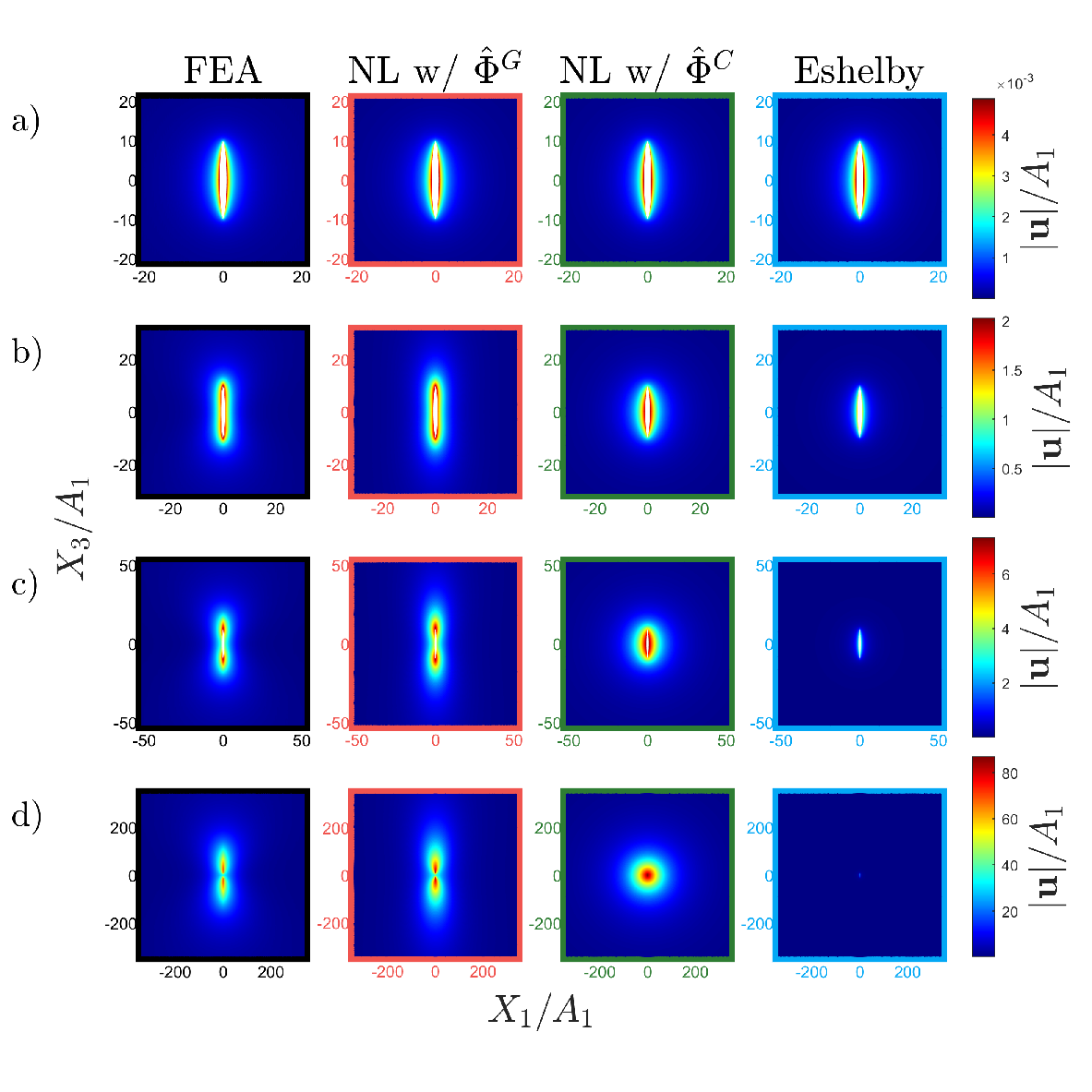}
    \caption{Full field plots of the normalized displacement magnitude $|\mathbf{u}|/A_1$ in the normalized reference coordinates $\X/A_1$ of an incompressible, neo-Hookean, isotropically growing prolate spheroidal inclusion-matrix system with undeformed aspect ratio $\Phi_0 = 10$. The volume growth ratios in each set of subplots are are (a) $J^*_I = 1.01$, (b) $J^*_I = 10$, (c) $J^*_I = 100$, and (d) $J^*_I = 12531$. Note that the colorbar scale and zoom level vary between each set of plots.}
    \label{fig:10_fullfield}
\end{figure}
\begin{figure}[t]
    \centering
    \includegraphics[width=0.98\textwidth]{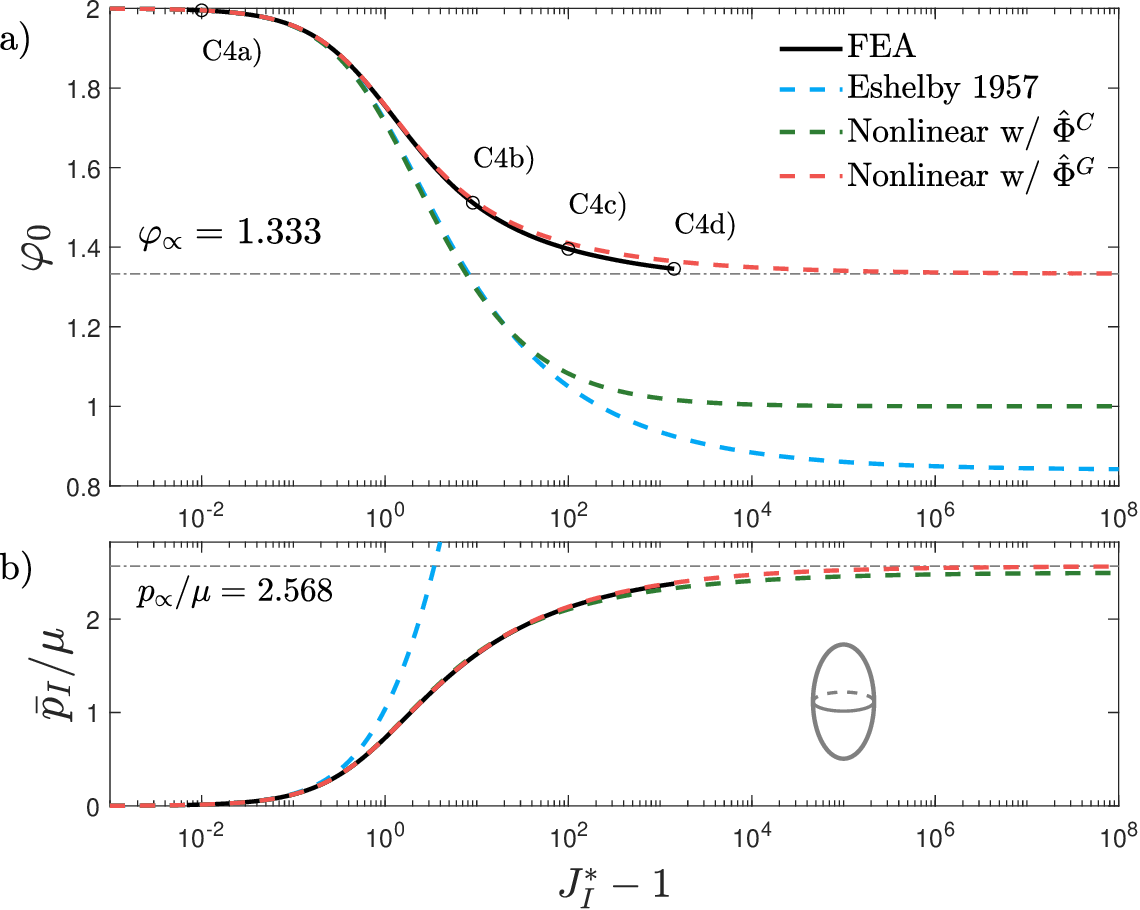}
    \caption{The deformed aspect ratio of the inclusion $\varphi_0$ (a) and volume averaged dimensionless inclusion pressure  $\bar{p}_I/\mu$ (b) vs. the volume growth ratio ($J^*_I-1$) of an incompressible and , neo-Hookean, isotropically growing prolate spheroidal inclusion-matrix system with undeformed aspect ratio $\Phi_0 = 2$. The dashed grey lines in (a) and (b) denote the isomorphic aspect ratio of the inclusion-matrix system $\varphi_\propto  = 1.333$ and the dimensionless isomorphic inclusion pressure of the inclusion-matrix system $p_\propto/\mu= 2.568$ respectively (as derived in Section \ref{sec:4.3}). The numerical labels in (a) denote the volume growth ratios of the deformation magnitude plots in Fig. \ref{fig:2_fullfield}.}
    \label{fig:2_2D}
\end{figure}
\begin{figure}[t]
    \centering
    \includegraphics[width=0.9\textwidth]{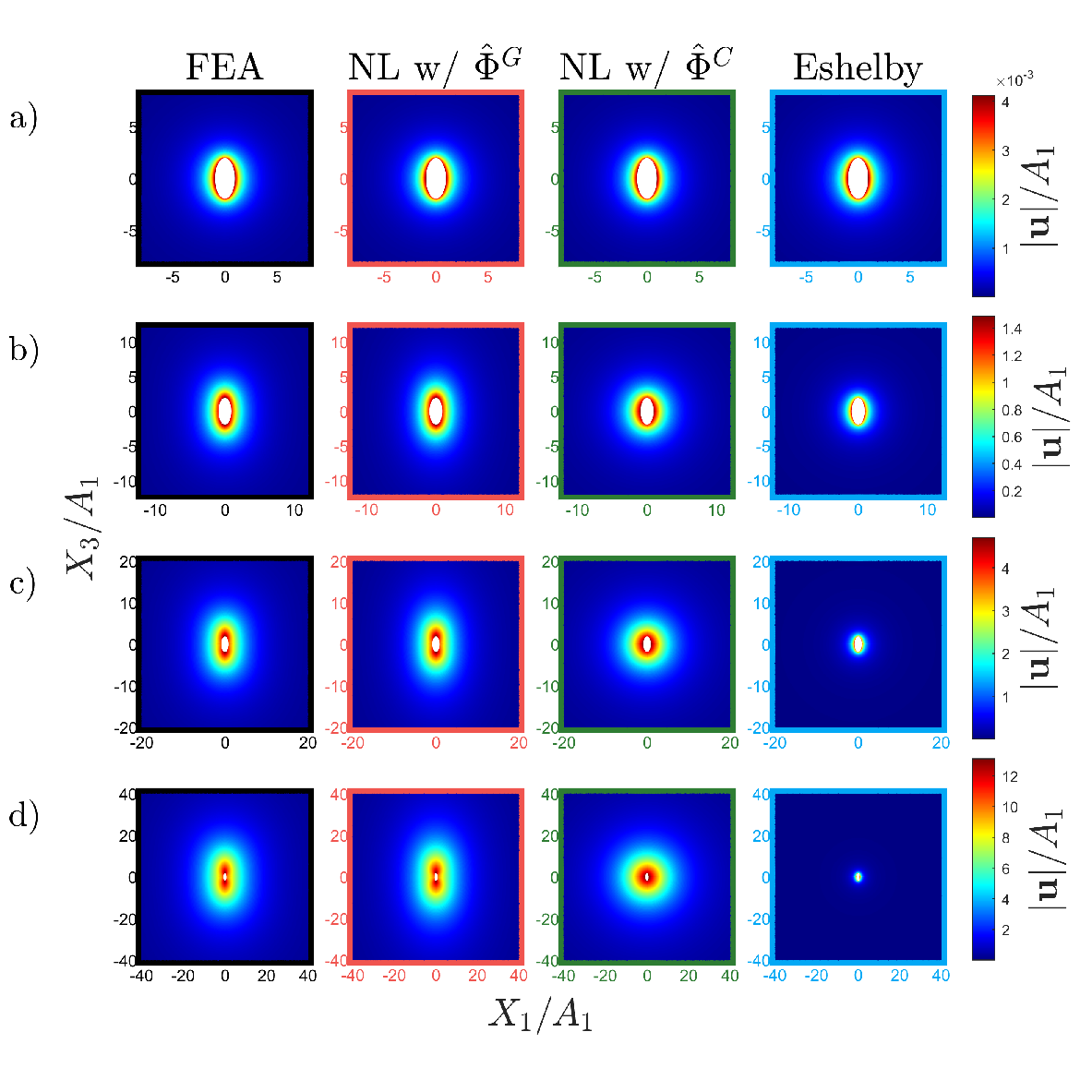}
    \caption{Full field plots of the normalized displacement magnitude $|\mathbf{u}|/A_1$ in the normalized reference coordinates $\X/A_1$ of an incompressible, neo-Hookean, isotropically growing prolate spheroidal inclusion-matrix system with undeformed aspect ratio $\Phi_0 = 2$. The volume growth ratios in each set of subplots are are (a) $J^*_I = 1.01$, (b) $J^*_I = 10$, (c) $J^*_I = 100$, and (d) $J^*_I = 55335$. Note that the colorbar scale and zoom level vary between each set of plots.}
    \label{fig:2_fullfield}
\end{figure}
\begin{figure}[t]
    \centering
    \includegraphics[width=0.98\textwidth]{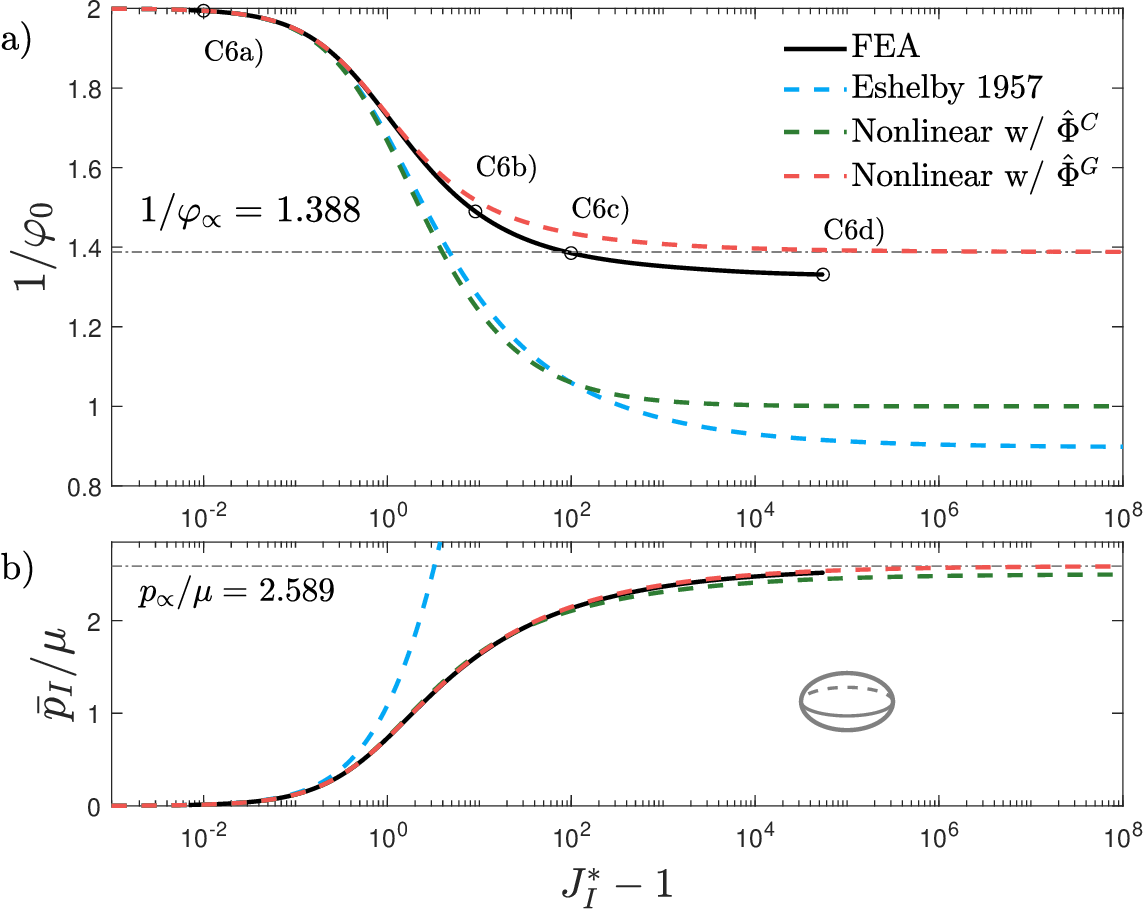}
    \caption{The deformed aspect ratio of the inclusion $\varphi_0$ (a) and volume averaged dimensionless inclusion pressure  $\bar{p}_I/\mu$ (b) vs. the volume growth ratio ($J^*_I-1$) of an incompressible and , neo-Hookean, isotropically growing prolate spheroidal inclusion-matrix system with undeformed aspect ratio $\Phi_0 = 1/2$. The dashed grey lines in (a) and (b) denote the isomorphic aspect ratio of the inclusion-matrix system $\varphi_\propto  = 0.720$ and the dimensionless isomorphic inclusion pressure of the inclusion-matrix system $p_\propto/\mu= 2.589$ respectively (as derived in Section \ref{sec:4.3}). The numerical labels in (a) denote the volume growth ratios of the deformation magnitude plots in Fig. \ref{fig:0.5_fullfield}.}
    \label{fig:0.5_2D}
\end{figure}
\begin{figure}[t]
    \centering
    \includegraphics[width=0.9\textwidth]{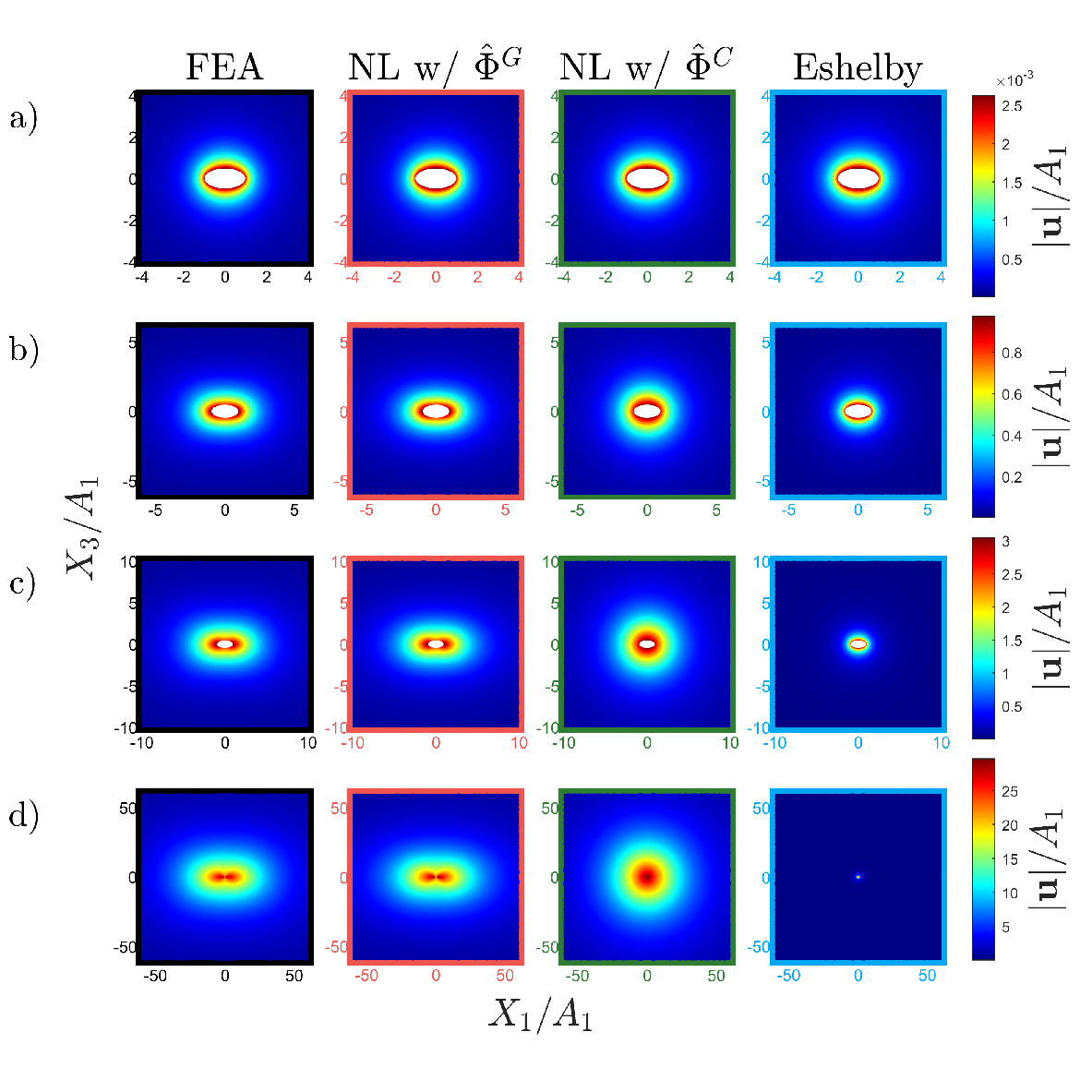}
    \caption{Full field plots of the normalized displacement magnitude $|\mathbf{u}|/A_1$ in the normalized reference coordinates $\X/A_1$ of an incompressible, neo-Hookean, isotropically growing prolate spheroidal inclusion-matrix system with undeformed aspect ratio $\Phi_0 = 1/2$. The volume growth ratios in each set of subplots are are (a) $J^*_I = 1.01$, (b) $J^*_I = 10$, (c) $J^*_I = 100$, and (d) $J^*_I = 1432$. Note that the colorbar scale and zoom level vary between each set of plots.}
    \label{fig:0.5_fullfield}
\end{figure}
\begin{figure}[t]
    \centering
    \includegraphics[width=0.98\textwidth]{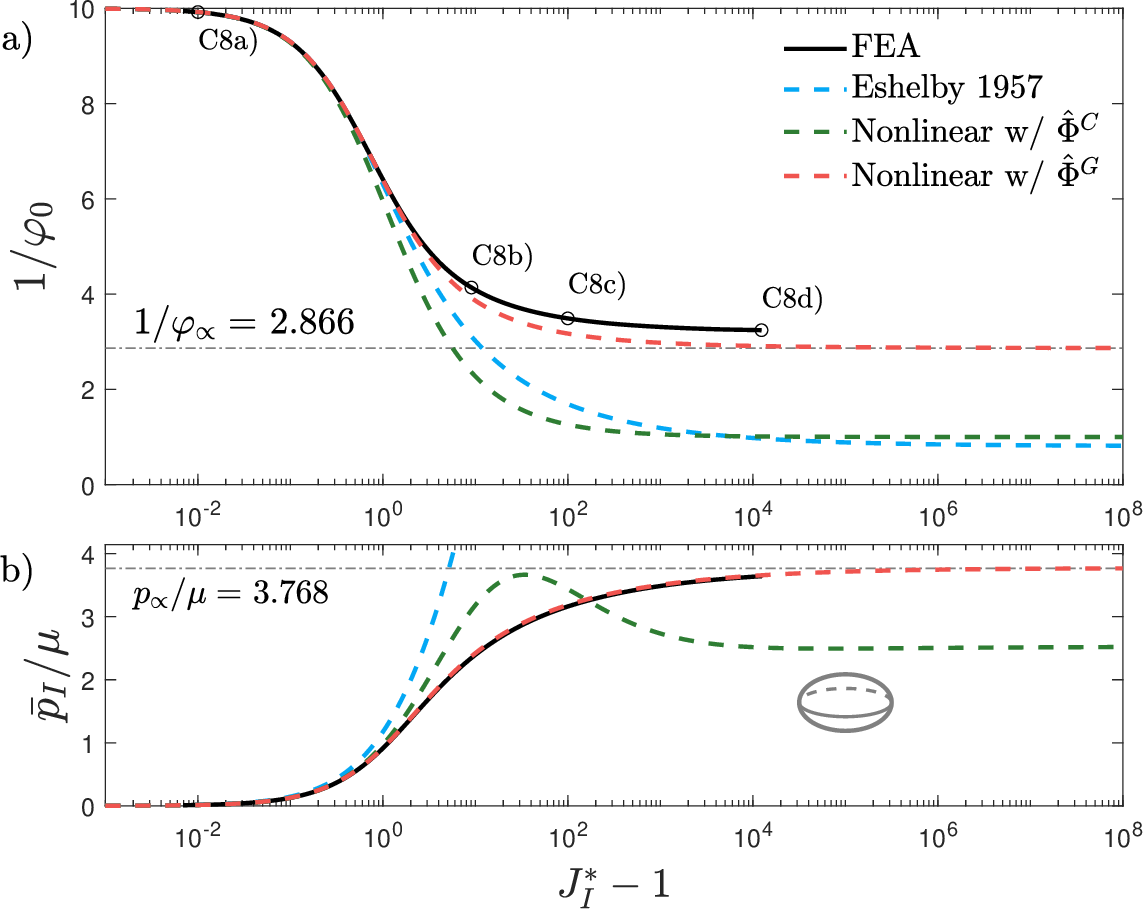}
    \caption{The deformed aspect ratio of the inclusion $\varphi_0$ (a) and volume averaged dimensionless inclusion pressure  $\bar{p}_I/\mu$ (b) vs. the volume growth ratio ($J^*_I-1$) of an incompressible and , neo-Hookean, isotropically growing prolate spheroidal inclusion-matrix system with undeformed aspect ratio $\Phi_0 = 1/10$. The dashed grey lines in (a) and (b) denote the isomorphic aspect ratio of the inclusion-matrix system $\varphi_\propto  = 0.349$ and the dimensionless isomorphic inclusion pressure of the inclusion-matrix system $p_\propto/\mu= 3.768$ respectively (as derived in Section \ref{sec:4.3}). The numerical labels in (a) denote the volume growth ratios of the deformation magnitude plots in Fig. \ref{fig:0.1_fullfield}.}
    \label{fig:0.1_2D}
\end{figure}
\begin{figure}[t]
    \centering
    \includegraphics[width=0.9\textwidth]{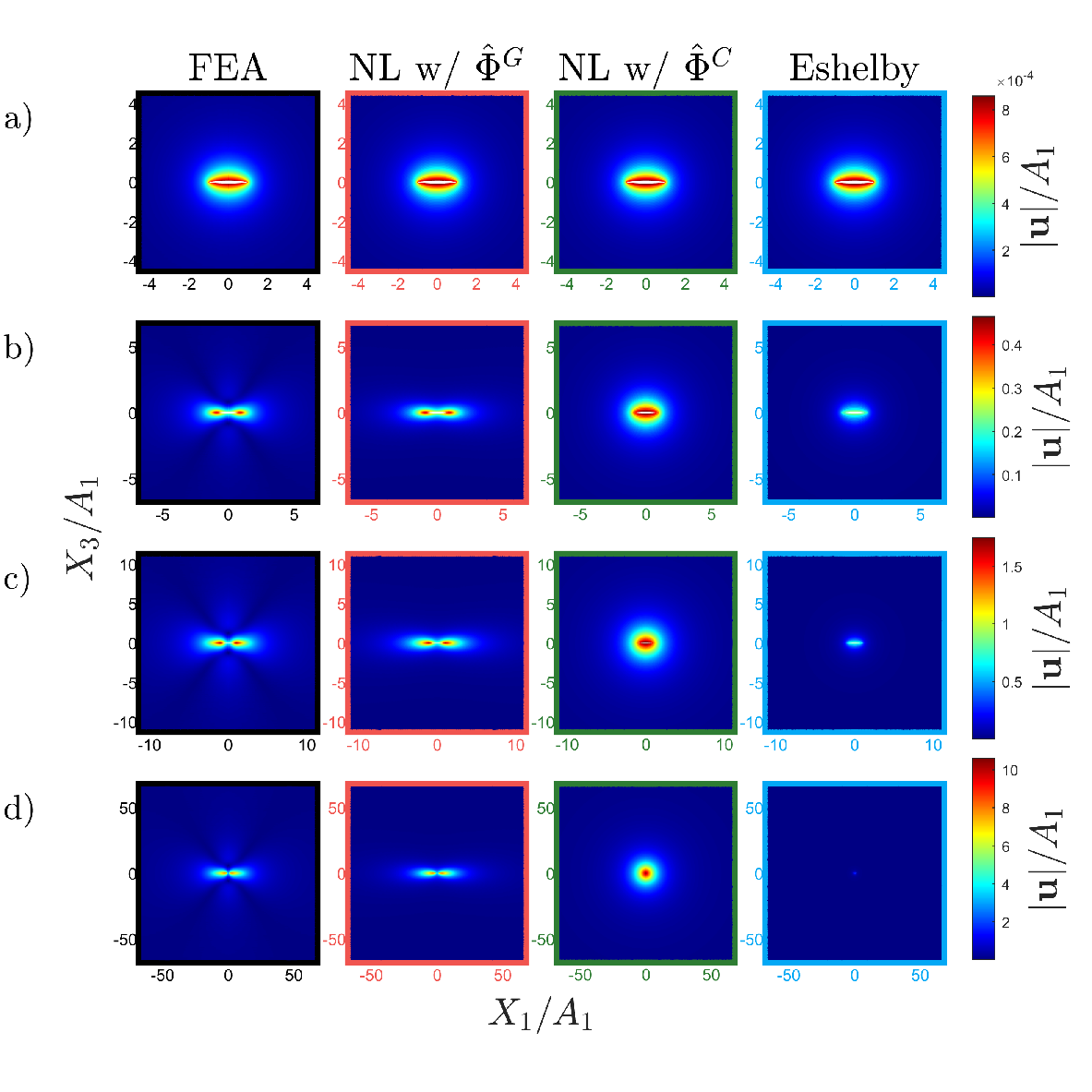}
    \caption{Full field plots of the normalized displacement magnitude $|\mathbf{u}|/A_1$ in the normalized reference coordinates $\X/A_1$ of an incompressible, neo-Hookean, isotropically growing prolate spheroidal inclusion-matrix system with undeformed aspect ratio $\Phi_0 = 1/10$. The volume growth ratios in each set of subplots are are (a) $J^*_I = 1.01$, (b) $J^*_I = 10$, (c) $J^*_I = 100$, and (d) $J^*_I = 69023$. Note that the colorbar scale and zoom level vary between each set of plots.}
    \label{fig:0.1_fullfield}
\end{figure}

\end{document}